\documentclass[usenatbib]{mn2e}
\usepackage{amsmath}
\usepackage{amssymb}
\usepackage[pdftex]{graphicx} 
\usepackage[pdftex,linktocpage]{hyperref} 
\newcommand{\beq}{\begin{eqnarray}} 
\newcommand{\eeq}{\end{eqnarray}} 
 
\newcommand{\apj}{ApJ} 
\newcommand{\apjs}{ApJS}

\newcommand{\mnras}{MNRAS} 
\newcommand{\aap}{A\&A}

\newcommand{\nat}{Nature}

\newcommand{\hMpc}{{\ifmmode{h^{-1}{\rm Mpc}}\else{$h^{-1}$Mpc }\fi}} 
\newcommand{\hGpc}{{\ifmmode{h^{-1}{\rm Gpc}}\else{$h^{-1}$Gpc }\fi}} 
\newcommand{\hmpc}{{\ifmmode{h^{-1}{\rm Mpc}}\else{$h^{-1}$Mpc }\fi}} 
\newcommand{\hkpc}{{\ifmmode{h^{-1}{\rm kpc}}\else{$h^{-1}$kpc }\fi}} 
\newcommand{\hMsun}{{\ifmmode{h^{-1}{\rm {M_{\odot}}}}\else{$h^{-1}{\rm{M_{\odot}}}$}\fi}} 
\newcommand{\hmsun}{{\ifmmode{h^{-1}{\rm {M_{\odot}}}}\else{$h^{-1}{\rm{M_{\odot}}}$}\fi}} 
\newcommand{\Msun}{{\ifmmode{{\rm {M_{\odot}}}}\else{${\rm{M_{\odot}}}$}\fi}} 
\newcommand{\msun}{{\ifmmode{{\rm {M_{\odot}}}}\else{${\rm{M_{\odot}}}$}\fi}} 
 
\begin{document} 
 
\title{A Dynamical Classification of the Cosmic Web}
\author[Forero-Romero et al.]
{\parbox[t]\textwidth{J.E. Forero-Romero $^{1}$, Y. Hoffman $^2$, S. Gottl\"ober $^1$,
  A. Klypin $^3$ \& G. Yepes $^4$}
\vspace*{6pt} \\
 $^{1}$Astrophysikalisches Institut Potsdam, An der Sternwarte 16, D-14482
  Potsdam, Germany\\ 
  $^{2}$Racah Institute of Physics, Hebrew University, Jerusalem 91904,
  Israel\\
  $^{3}$Department of Astronomy, New Mexico State University, Box 30001,
  Department 4500, Las Cruces, NM 880003, USA\\
  $^{4}$Grupo de Astrof\'{\i}sica, Universidad Aut\'onoma de Madrid,
  Madrid E-28049, Spain \\
}
\date{\today} 
 
\maketitle

\begin{abstract}

A dynamical classification of the cosmic web is proposed. The  large scale environment
is classified into four web types: voids, sheets, filaments and knots. The
classification is based on the evaluation of the deformation tensor, i.e. the
Hessian of the gravitational potential, on a grid. The classification is based
on counting the number of eigenvalues above a certain threshold,
$\lambda_\mathrm{th}$, at each grid point, where the case of zero, one, two or
three such eigenvalues corresponds to void, sheet, filament or a knot grid
point. The collection of neighboring grid points, friends-of-friends, of the
same web attribute   constitutes  voids, sheets, filaments and knots as web
objects. 

A simple dynamical consideration suggests that $\lambda_\mathrm{th}$ should be
approximately unity, upon an appropriate scaling of the deformation tensor.  The
algorithm has been applied and tested against a suite of (dark matter only)
 cosmological N-body simulations. In particular, the dependence of the volume and mass
filling fractions on $\lambda_\mathrm{th}$ and on the resolution has been
calculated for the four web types. Also, the   percolation properties of voids
and filaments have been studied.

Our main findings are:
(a) Already at $\lambda_\mathrm{th}=0.1$ the resulting web classification
reproduces  the visual impression of the cosmic web. 
(b)  Between $0.2 \lesssim \lambda_\mathrm{th} \lesssim 0.4 $, a system of percolated voids
coexists with a net of interconected filaments. This suggests a reasonable
choice for $\lambda_\mathrm{th}$ as the parameter that defines the cosmic
web. 
(c) The dynamical nature of the suggested classification provides a robust
framework for incorporating environmental information into galaxy formation
models, and in particular the semi-analytical ones.

\end{abstract}

\section{Introduction} 
\label{sec:intro} 
 
The large scale structure of the Universe, as depicted from galaxy surveys, weak lensing mapping and 
numerical simulations, shows a web like  three dimensional 
structure. There are three features that can be generally 
observed. First,  most of the volume resides in underdense regions; 
second, most of the volume is permeated by filaments; third, the 
densest clumps are located at the intersection of filaments 
\citep{1996Natur.380..603B}.  This motivates a classification of the cosmic 
web into at least three categories: voids (underdense regions), filaments and 
knots (densest clumps). 
 
There are  clear evidences for the  correlations of the observed  properties of galaxies with the 
environment.  We have for instance the morphology density correlation,  where
elliptical galaxies are found preferentially in crowded  environments, and spiral galaxies are found in the field 
\citep{1980ApJ...236..351D}. The same kind of correlation can be found in
terms of the colors of the galaxies \citep{2005ApJ...629..143B}.

According to the current paradigm of  structure formation  galaxies 
form and evolve in  dark matter (DM) halos \citep{1978MNRAS.183..341W}. It
follows that the study of such environmental dependence should commence  with
the effort to understand the formation of DM halos in the context of the
cosmic web
\citep{2005MNRAS.363L..66G,2005ApJ...634...51A,2007ApJ...654...53M}. This
motivates us to search for a robust and meaningful  method   to classify  the
different environments in numerical simulations. Such a classification should
provide the framework for studying  the environmental dependence of  galaxy formation.

 Translating the visual impression into an algorithm that classifies the local 
 geometry into different environments is not a trivial task. A somewhat less 
 challenging, yet very closely related, task is that of identifying just the 
 voids out of the cosmic web. A thorough review and comparison of different 
 algorithms of void finders  has been recently presented in 
 \cite{2008MNRAS.387..933C}.  The   void finders can be classified according 
 to the method employed. Most are based on the point distribution of galaxies 
 or dark matter (DM) halos and some on the smoothed density or potential 
 fields. Some of the finders are based on spherical filters  while others 
 assume no inherent symmetry
 \citep{2007MNRAS.375..184B,2005MNRAS.360..216C,2003MNRAS.344..715G, 
   2002MNRAS.332..205A, 2008MNRAS.386.2101N, 2007MNRAS.380..551P,  
   2002MNRAS.330..399P, 2006MNRAS.367.1629S}.  
 
 It should be emphasized that an environment finder should be evaluated by its merits 
 and it cannot be labeled as correct or wrong. A good algorithm should provide 
 a quantitative classification which agrees with the visual impression and it 
 should be based on a robust and well defined numerical scheme. Yet, it is 
 desirable for an algorithm to be based on an analytical prescription so that 
 its outcome can be calculated analytically. Also, simplicity is always  very 
 highly desired.

A variety of approaches have been employed in the classification of the cosmic 
environment into its basic elements. The simplest way is based on the 
association of the environment with the local density, evaluated with a 
top-hat filter, say, of some width \citep{1999MNRAS.302..111L}.  The density 
field can be analyzed in a much more sophisticated and elaborated way. This is 
the case of the web classification based on the multi-scale analysis of the 
Hessian matrix of the density field  \citep{2007A&A...474..315A} 
or the skeleton analysis of the density field \citep{2006MNRAS.366.1201N,2008MNRAS.383.1655S}. Both
methods classify the cosmic web by pure geometrical tools applied to the
density field. A very different approach is done within a dynamical framework
in which the analysis of the gravitational potential is used to classify the
web. This has been inspired by the seminal work of \cite{1970A&A.....5...84Z}
that led  to the "Russian school of structure formation"
(e.g. \cite{1982GApFD..20..111A, 1983MNRAS.204..891K}). The quasi-linear
theory of  the Zeldovich approximation predicts the existence of an
infinitely  connected web of pancakes (i.e. sheets), filaments and knots. This
morphological classification is based on the study of the eigenvalues of the
deformation tensor, namely the Hessian matrix of the linear gravitational field.

A recent application of the Zeldovich-based classification has been provided 
by \cite{2008arXiv0801.1558L} who used a Wiener filter linear reconstruction of the 
local density field and evaluated the linear deformation, and hence also the shear, 
tensor on a grid. The cosmic web has been classified according the structure 
of the shear tensor.  
 
A different approach has been followed by \cite{2007MNRAS.375..489H}   who
suggested that the full non-linear  gravitational potential should be used for
the  geometrical classification. Apart from the difference between the linear
and the  non-linear potential, both \cite{2007MNRAS.375..489H} and
\cite{2008arXiv0801.1558L}  are using the same classification. Namely, the
Hessian of the gravitational  potential is evaluated on a grid and its
eigenvalues are examined locally. A  grid point is classified as a void, sheet,
filament or knot point if the  number of eigenvalues greater that a null
threshold is zero, one, two or three.

The algorithm presented here is an extension and improvement of the one 
suggested by \cite{2007MNRAS.375..489H}. The extension is represented in the
selection of a new free parameter with a dynamical interpretation. As such it provides a  
classification of local environment. Namely each spatial point is flagged as
belonging to a either a void, sheet, filament or a knot point. It is the collective 
classification of all points in space which gives rise to the geometrical 
construction we call the cosmic web. This opens the door for defining voids, 
sheets, filaments and knots as individual objects. Each object is defined as a
collecion of connected points having the same environmental attribute. We use
a friends-of-friends (FoF) algorithm to find connected sets of points in
simulations. Having defined the web objects the 
statistical and dynamical properties of these can be readily studied. Here we 
shall focus on analyzing the statistical properties of the voids and
filaments, aiming to better constrain the value for the free parameter we
introduce. We perform as well a statistical study to asses effect of the
cosmic variance on our conclusions.

This paper is organized as follows. 
The web classification scheme is described in \S \ref{sec:web}. 
The N-body simulation used in the paper and the numerical implementation of
the web classification are presented in \S \ref{sec:n-body}.  \S
\ref{sec:res-web} describes the main properties of the cosmic web and in
particular its dependence on the smoothing scale and the free parameter of our classification
scheme. \S  \ref{sec:res-void} concentrates on the properties of the voids
sector of the cosmic web. \S \ref{sec:frag-filament} studies the fragmentation
of filaments in order to give a confidence interval to the free parameter that
was introduced. In \S \ref{sec:cosmic-variance} we revisit the sections \S
\ref{sec:res-web} and \S \ref{sec:res-void} to study the effect of cosmic
variance. The paper concludes with a general discussion and a summary of the
main results of the paper  (\S \ref{sec:disc}).

\section{Web Classification} 
\label{sec:web} 
 
 \cite{2007MNRAS.375..489H}  have recently suggested a new dynamical 
 classification of the cosmic web. The basic idea of their approach is that 
 the eigenvalues of the deformation tensor determine the geometrical  nature of each 
 point in space. 
 
 The deformation tensor,  $T_{\alpha\beta}$, is defined by the Hessian  of 
 the gravitational potential $\phi$:  
 \begin{equation}
\label{eq:hessian}
 T_{\alpha\beta}= {\partial ^2 \phi \over \partial r_\alpha r_\beta}.
\end{equation}
 
The definition of the deformation tensor \footnote{
 \cite{2007MNRAS.375..489H} call the $T_{\alpha\beta}$ tensor the 'tidal tensor'. 
Usually the tidal tensor is defined as the traceless part of $
 T_{\alpha\beta}$.} explicitly assumes that the matter density 
field is known and that it is smoothed with a finite kernel, or otherwise the 
derivatives are not defined. For simplicity the smoothed density field is 
defined over a (Cartesian) grid.  
 
\cite{2007MNRAS.375..489H}  considered the three eigenvalues of the deformation
tensor, $\tilde{\lambda}_1\ \ge\ \tilde{\lambda}_2 \ \ge \ \tilde{\lambda}_3$, and classified a grid point according the number of 
positive eigenvalues at that point.  Namely, a void point corresponds to no 
positive eigenvalues, a sheet to one, a filament to two and a knot point to 
three positive eigenvalues. The sign of a given eigenvalue at a given grid 
node determines whether the gravitational force at the direction of the 
principal direction of the corresponding eigenvector  is contracting (positive 
eigenvalue) or expanding (negative).

\cite{2007MNRAS.375..489H}  provided a very attractive approach to the web 
classification problem. It is based on the dynamical nature of the web, and so 
it easily lends itself to a theoretical analysis. The ease of its application 
to cosmological simulations opens the door for a new framework for associating 
the properties of galaxies and dark matter  with environment, as defined 
by the web classification.  
 
Close inspection of the \cite{2007MNRAS.375..489H} classification scheme and 
its results reveals its shortcomings. The volume filling factor  of 
\cite{2007MNRAS.375..489H} voids is very small. For their minimal smoothing 
scale, namely the highest resolution, the voids occupy only $17\%$ of the 
simulated volume. This stands in contrast to  the visual 
impression of voids in the actual universe and in simulation, where voids 
seems to occupy most of the volume but contain only a small fraction of the 
galaxies (in observations) or matter (in simulations).  Furthermore, the 
\cite{2007MNRAS.375..489H} classification does not reproduce the visual 
perception of the cosmic web.  
 
It is easy to understand the inability of the \cite{2007MNRAS.375..489H} 
approach to reproduce the visual impression. The web classification is based 
on the algebraic sign of the eigenvalues of the deformation tensor, namely the 
number of eigenvalues larger than a threshold value of zero. It follows that 
if an eigenvalue is only infinitesimally positive, the scheme assumes that the 
local neighborhood of the given grid point collapses along the 
corresponding eigenvector. Yet, the collapse proceeds over the dynamical time 
scale and if the value of the eigenvector is small enough the collapse will 
occur, if at all, only  in the distant future. Visual inspection would not 
classify the region as collapsing at the present time. This leads us to 
consider an alternative  approach, namely asking the eigenvalues to be larger
than a positive threshold.

Seeing that the dimensionality of the deformation tensor is  $[\mathrm{
  time}]^{-2}$, its eigenvalues  can be associated with the collapse time.  It
  follows that the threshold value should be roughly determined by equating
  the collapse time with the age of the universe. In Appendix
  \ref{app:setting} we rewrite the eigenvectors of the deformation tensor in a
  dimensionless way. In such a presentation  one expects $\lambda_\mathrm{th}
  \approx 1$,  where $\lambda_\mathrm{th}$ is the threshold value. For an
  isotropic collapse 
  $\lambda_\mathrm{th}$ can be calculated explicitly (see Eqs. (\ref{eq:tff}) -
  (\ref{eq:final})). However, this provides a very rough guide as the collapse on
  the web is clearly not isotropic. Here, an empirical approach is to be used
  and $\lambda_\mathrm{th}$ is to be roughly determined from the geometrical
  nature of the cosmic web it defines.

\begin{figure*} 
\begin{center} 
\includegraphics[width=6.8cm]{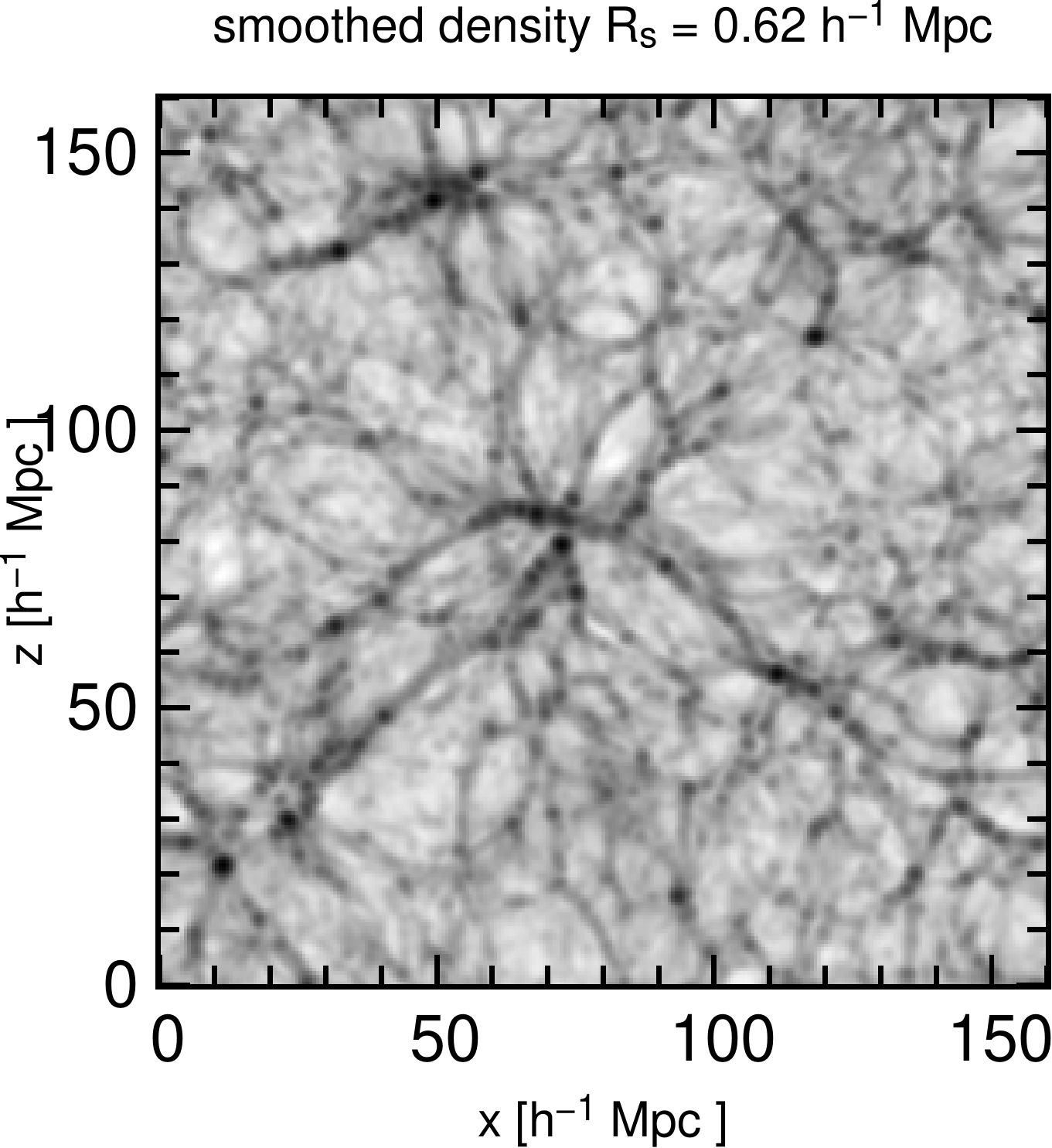}\hspace{0.5cm}
\includegraphics[width=6.8cm]{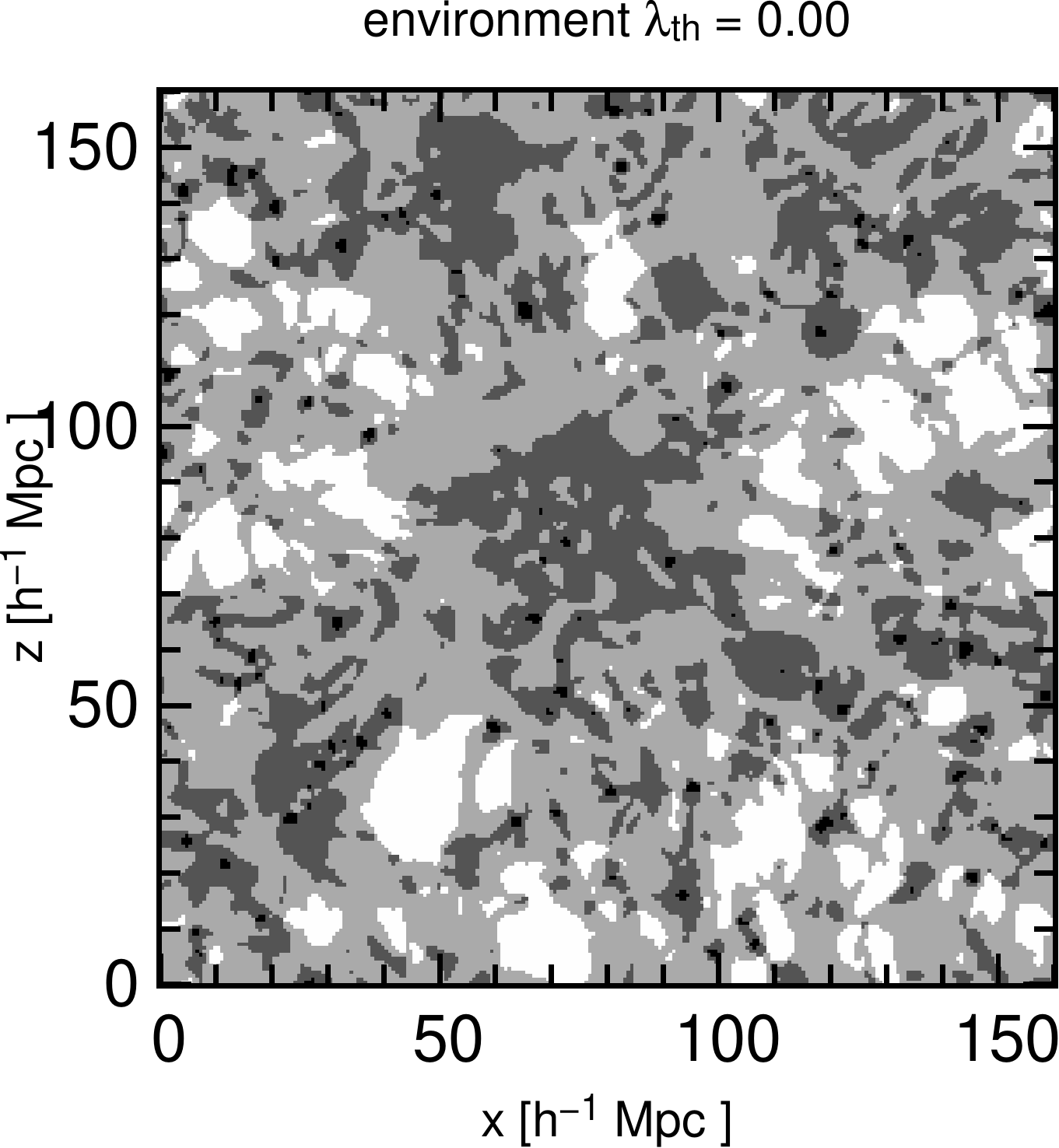} 
\includegraphics[width=6.8cm]{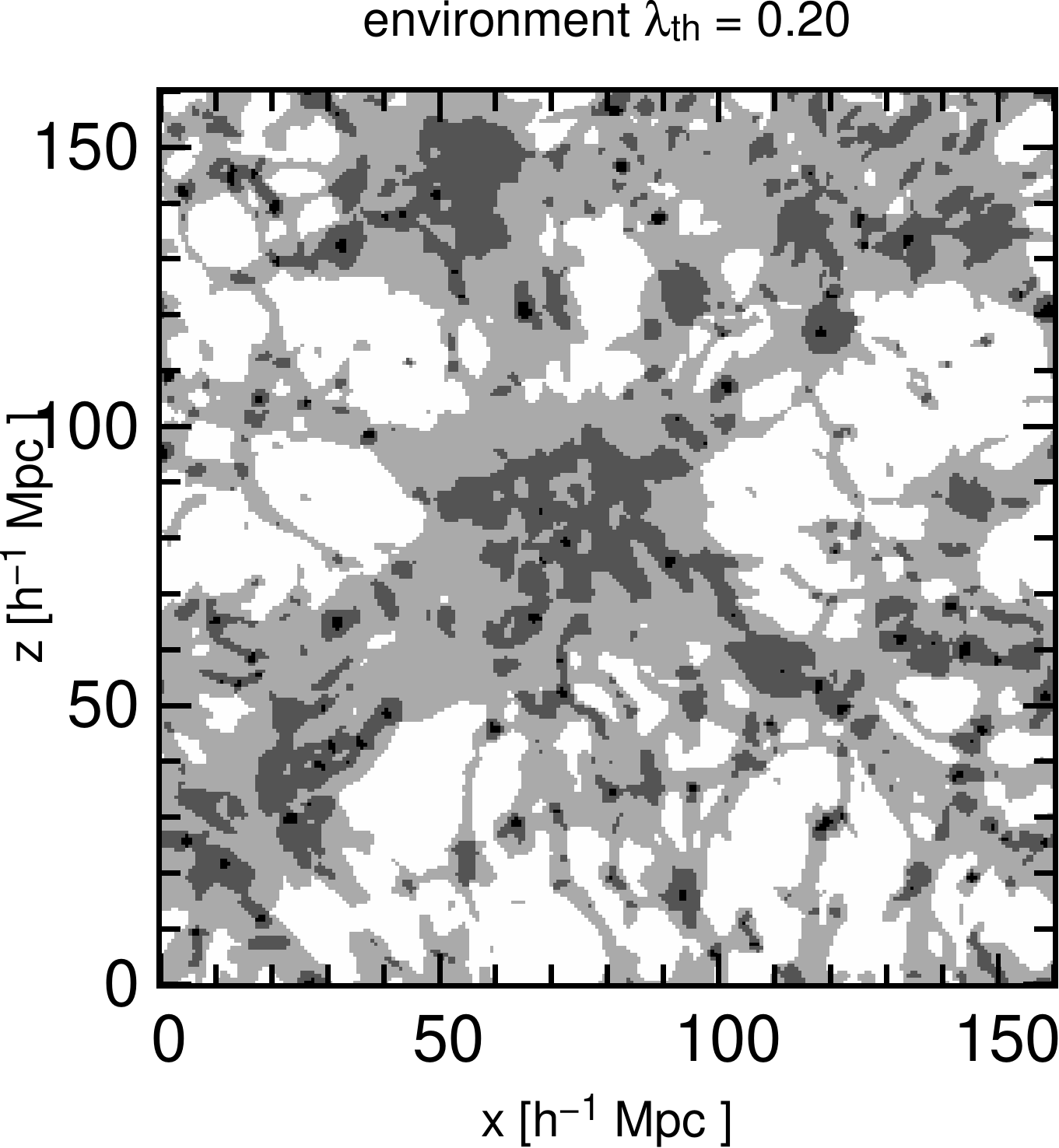}\hspace{0.5cm} 
\includegraphics[width=6.8cm]{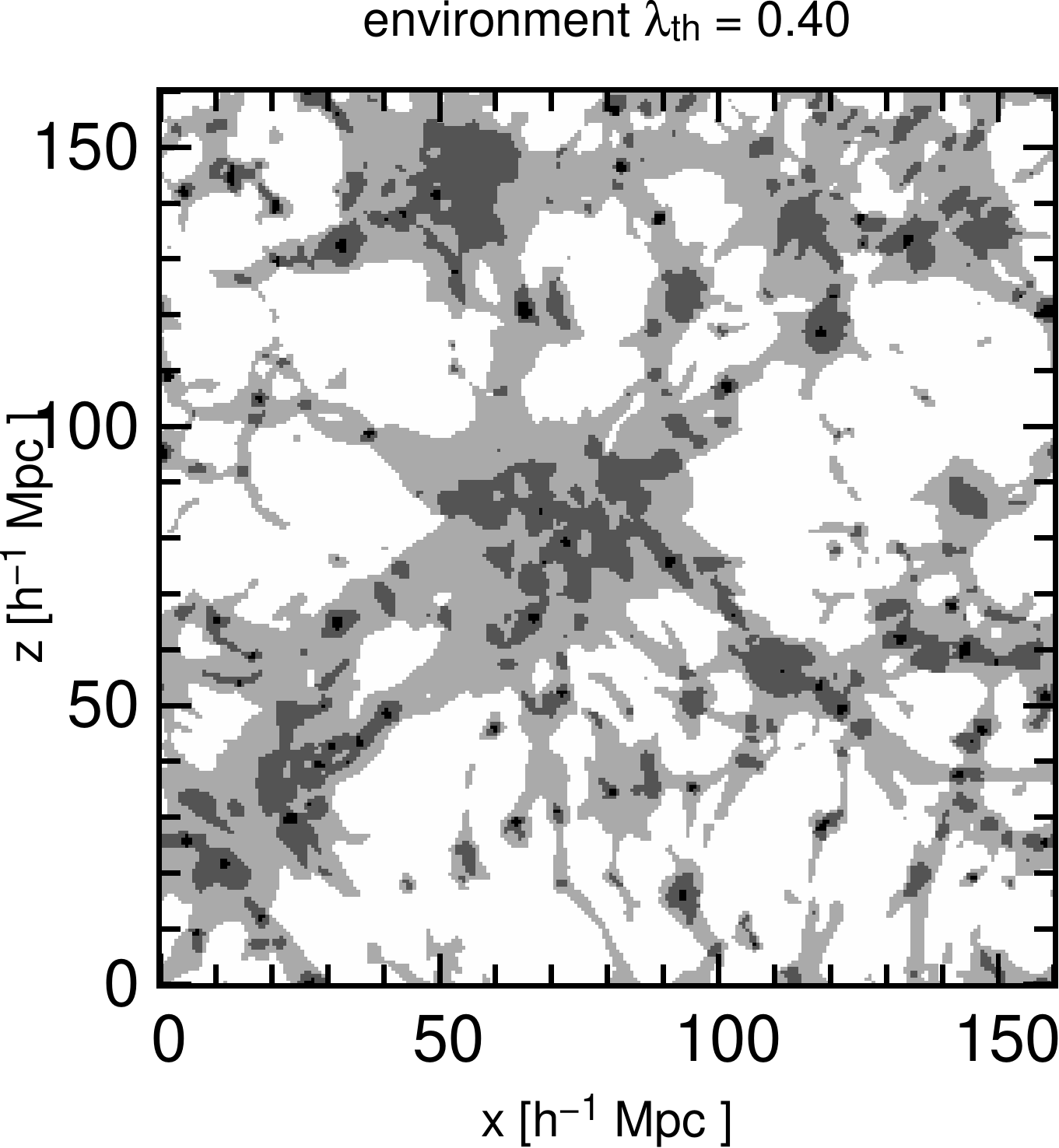} 
\includegraphics[width=6.8cm]{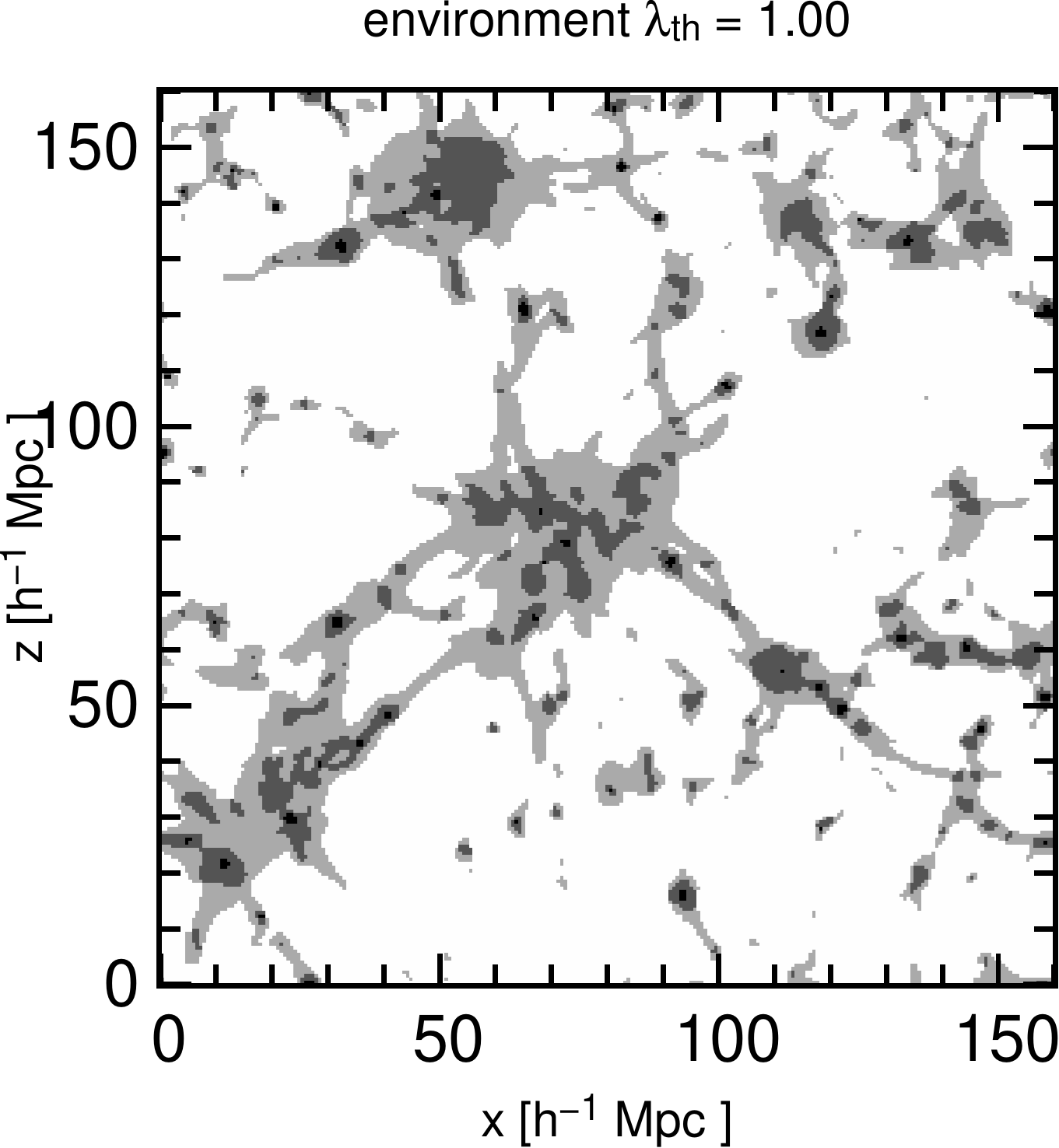}\hspace{0.5cm} 
\includegraphics[width=6.8cm]{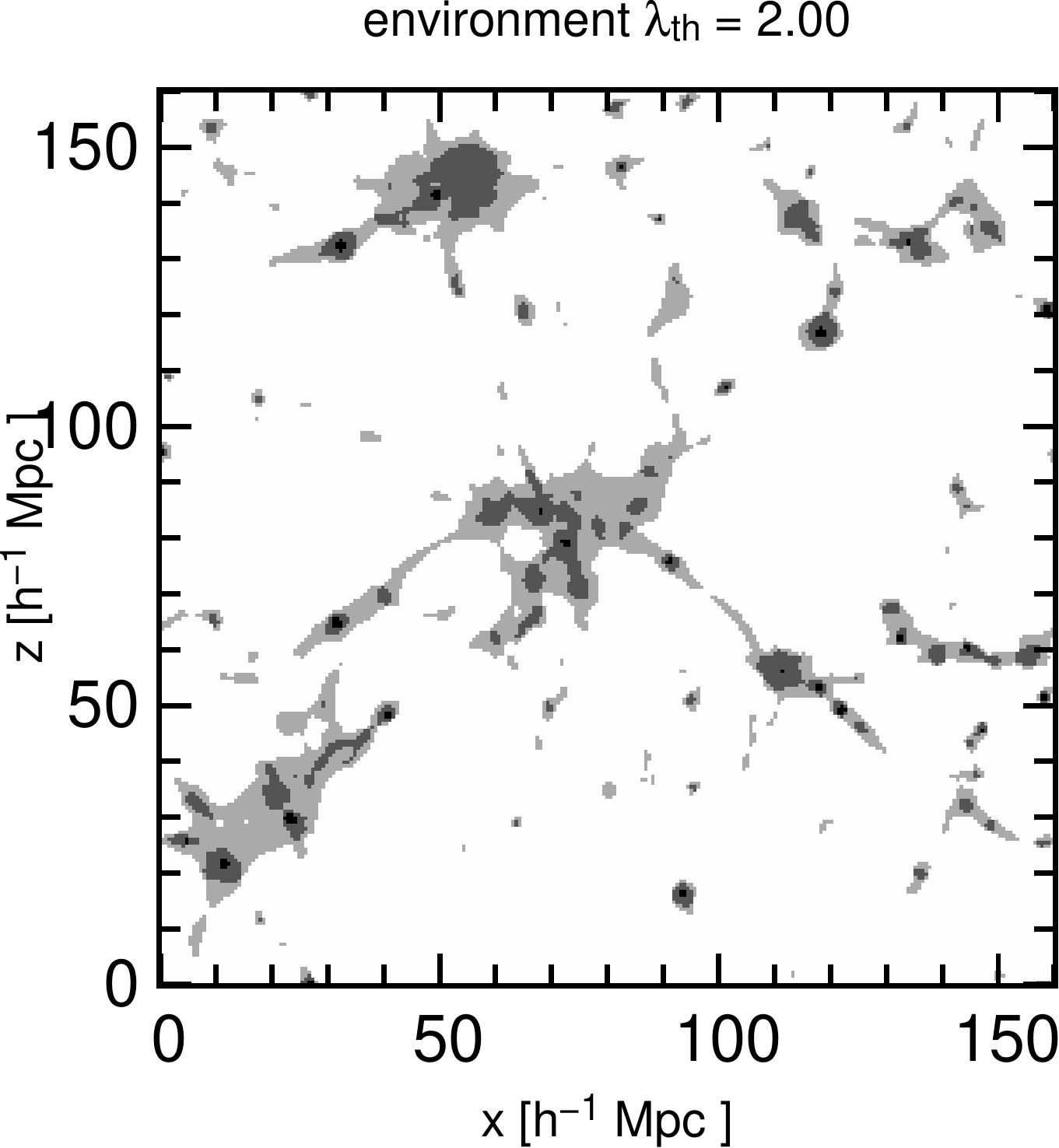} 
\caption{The density field and  five different environmental
  classifications. \emph{Upper left panel}: slice of width $0.625 \hMpc$
 depicting the density field in the simulation smoothed over a scale of
 $R_\mathrm{s}=0.625 \hMpc$, the color coding is
logarithmic in the density --- high density peaks are dark. The other
panels show the environment classification using different values in
for the threshold, $\lambda_\mathrm{th}=0.0, 0.20, 0.40, 1.0$, and $2.0$. 
White corresponds to voids, clear gray to
sheets, dark gray to filaments and black to peaks. The general
impression is that the non-zero values of $\lambda_\mathrm{th}$ below $1.0$
 capture better the environment seen by eye in the density plot.} 
\label{fig:dens_field}
\end{center}
\end{figure*}

\section{N-body simulation, Numerical Implementation and Object Detection} 
\label{sec:n-body}

We use two numerical simulation. The first assumes a WMAP3 cosmology 
\citep{WMAP3} with a matter density $\Omega_{\rm m} = 0.24$, a 
cosmological constant $\Omega_{\Lambda} = 0.76$, a dimensionless 
Hubble parameter $h = 0.73$, a spectral index of primordial density 
perturbations $n = 0.96$ and a normalization of $\sigma_{8} = 0.76$. 
A simulation of box size $160 \hMpc$ and $1024^3$ particles has been assumed, 
corresponding to a particle mass of $3.5 \times 10^8 \Msun$. Starting at 
redshift $z=30$   the evolution is followed using the MPI version of the Adaptive 
Refinement Tree (ART) code described in \cite{ART}. The simulation used here 
is actually a constrained simulation of the local universe which is to be 
described at length at the forthcoming Yepes  et al paper (in preparation). 
This is an updated and higher resolution version of the constrained simulation 
presented in    \cite{2003ApJ...596...19K}. Here the simulation is treated as just a 
random one and its constrained nature is completely ignored.

In order to estimate cosmic variance effects we have used a numerical
simulation of box size $1 \hGpc$. The assumed cosmology for this simulation
is WMAP3-like with a matter density $\Omega_{\rm } = 0.27$, a cosmological
constant $\Omega_{\Lambda} = 0.73$, a dimensionless
Hubble parameter $h = 0.70$, a spectral index $n=0.95$, a normalization
$\sigma_{8} = 0.79$ and $1024^3$ particles corresponding to a particle mass of
$9.8\times 10^{10} \Msun$. The simulation was performed using the ART code as
well.

The analysis of the simulations proceeds as follows. The density field  of the
$160\hMpc$ simulation is  calculated from the particle distribution on a $256^3$ grid using the 
Cloud-In-Cell (CIC) scheme, it is then smoothed with a Gaussian 
kernel of width $R_\mathrm{s}$ and from which the deformation tensor is calculated 
directly, using an FFT solver. The deformation tensor is then diagonalized on the 
grid. The web characteristic of each grid point is determined by the number of 
eigenvectors, at that grid node, above the threshold. It should be realized 
that the classification is local by its nature, but the combined effect of all 
grid points results in the geometrical construction defined as the cosmic 
web. 

In the $1\hGpc$ simulation we followed the same procedure, interpolating first the
density on a $512^3$ grid. Once the environment detection procedure is done we
select $6^3=216$ non-intersecting sub-boxes of $160\hMpc$ on a side to test
the effect of cosmic variance. Figure \ref{fig:dens_field} presents the CIC
density field and the cosmic web  evaluated at the  threshold values of
$\lambda_\mathrm{th}=0.00, \ 0.20, \ 0.40, \ 1.00$ and $2.00$. The web is
presented by a grey scale corresponding to the  four web types, it is
evaluated at a Gaussian smoothing of  $R_\mathrm{s} = 0.625 \hMpc$ in the
Figure \ref{fig:dens_field}. The  density field and the cosmic web are evaluated on a plane of  the CIC
grid. Visual inspection of the density field reveals a network of  voids,
filaments and dense knots. The density field is evaluated on a thin  plane and
therefore no clean distinction can be made between the 3-dimensional
filaments and sheets.

The web defined by  $\lambda_\mathrm{th}=0.0$ consists 
of many small  isolated voids that occupy only a small fraction of the total 
area of the plane.  Only as $\lambda_\mathrm{th}$ increases the voids get 
bigger and connected and they become the dominant geometrical component of the 
web.

The non-zero threshold classification provides a better visual match to the 
density field than the null case.  A qualitative analysis and comparison is presented in \S
\ref{sec:res-web}. The analysis is based on two quantities: the volume
occupied by each web type (volume filling fraction - VFF) and the fraction of
mass contained in such a volume (mass filling fraction - MFF).

A characterization of the cosmic web is obtained by grouping, by a
Friends-of-Friends (FoF) algorithm, neighboring grid points of a given web
type into individual objects. The resulting objects are defined as voids,
sheets, filaments and knots. The FoF  association proceeds in the following way: the centers
of the cells in the grid are used as the position of four different kinds of
particles according to its web type, then a standard FoF is run over particles
of the same kind with a linking length $b=1.1$ times the grid length. It means
that only the six closest neighbors of a given cell are taken into
account. The FoF void detection is done for different simulations: the $160 \hMpc$
simulation, the $1\hGpc$ simulation and 216 sub-boxes of $160 \hMpc$ on a side extracted
from the largest simulation. To be consistent in  the analysis of these three kinds
of simulations, one should make the FoF detection \emph{without} considering the
periodic boundary conditions, as it is the right boundary condition for the
sub-volumes extracted from the $1\hGpc$ simulation.

The FoF algorithm is used to detect voids for different
threshold values in the different simulations which are smoothed with
the same physical scale $R_\mathrm{s}=1.95\hMpc$ in order to allow a fair
comparison between them. In \S \ref{sec:res-void} a detailed analysis on these
voids is presented, paying special attention to its percolation properties as
the threshold rises.  The same kind of analysis is performed for the 
filaments in \S\ref{sec:frag-filament}, but only on the $1\hGpc$ simulation.

\section{Volume and Mass Filling Fractions} 
\label{sec:res-web}

\begin{figure*} 
\begin{center} 
\includegraphics[width=7cm]{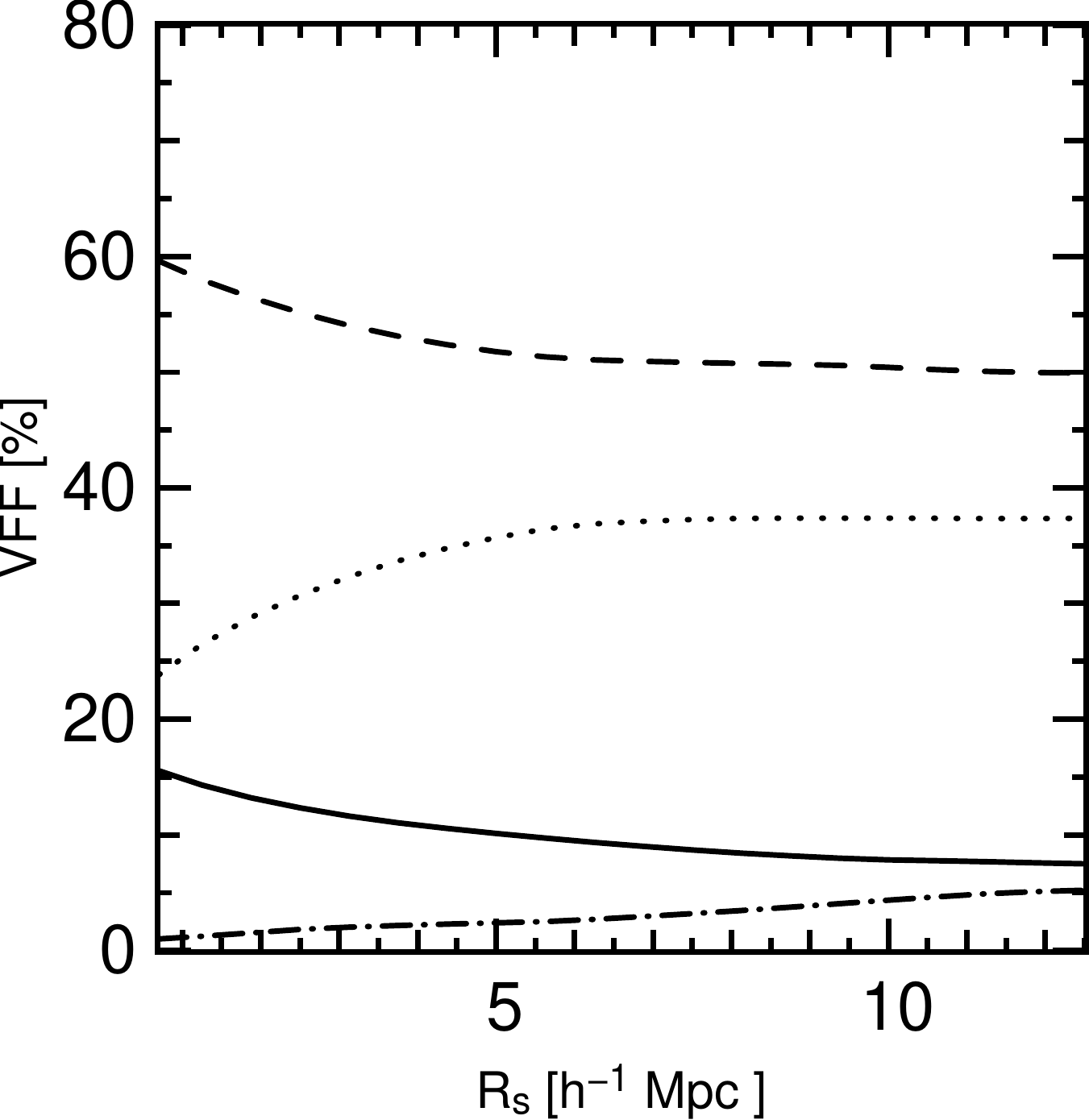}\hspace{0.5cm} 
\includegraphics[width=7cm]{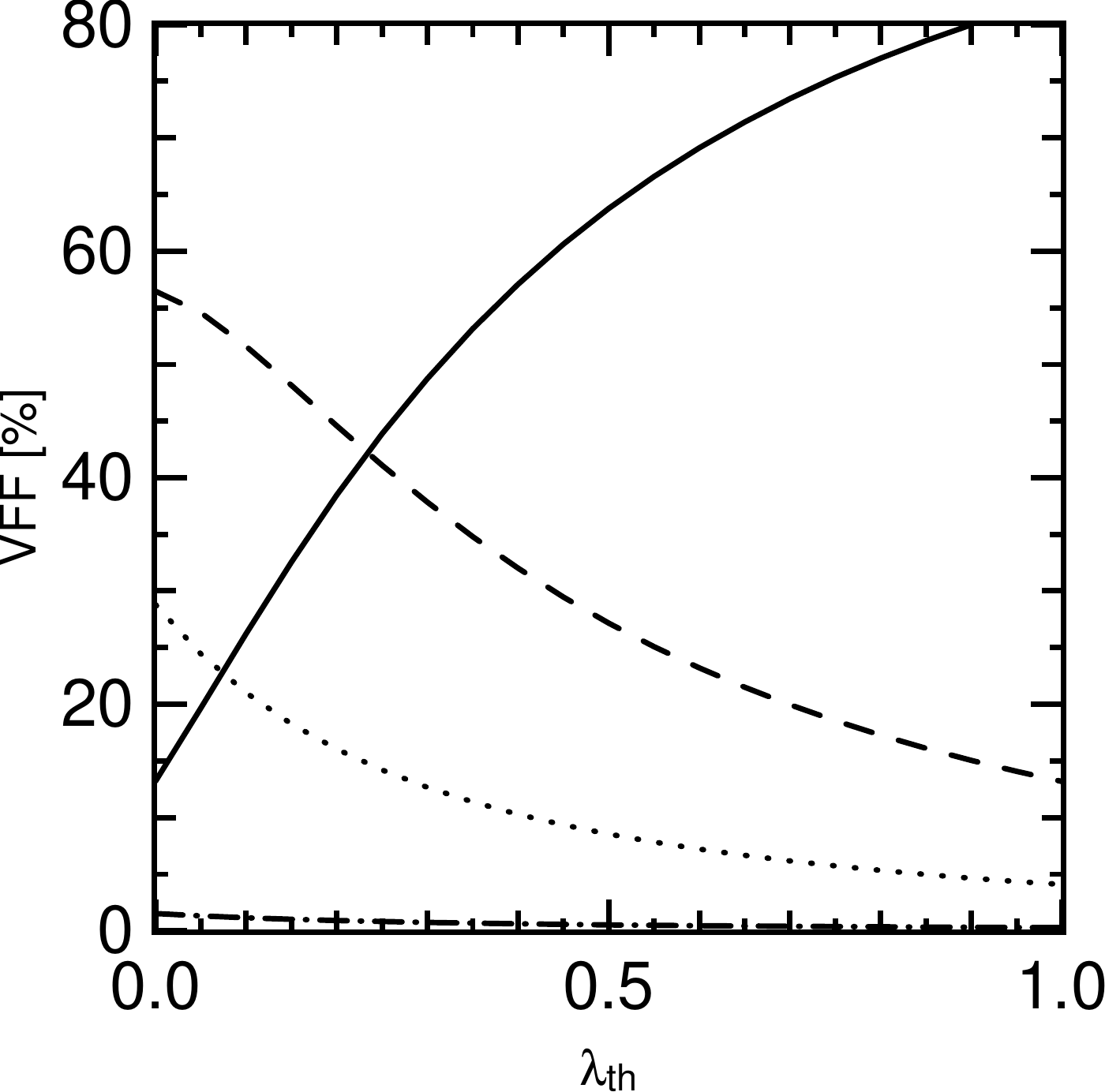}
\includegraphics[width=7cm]{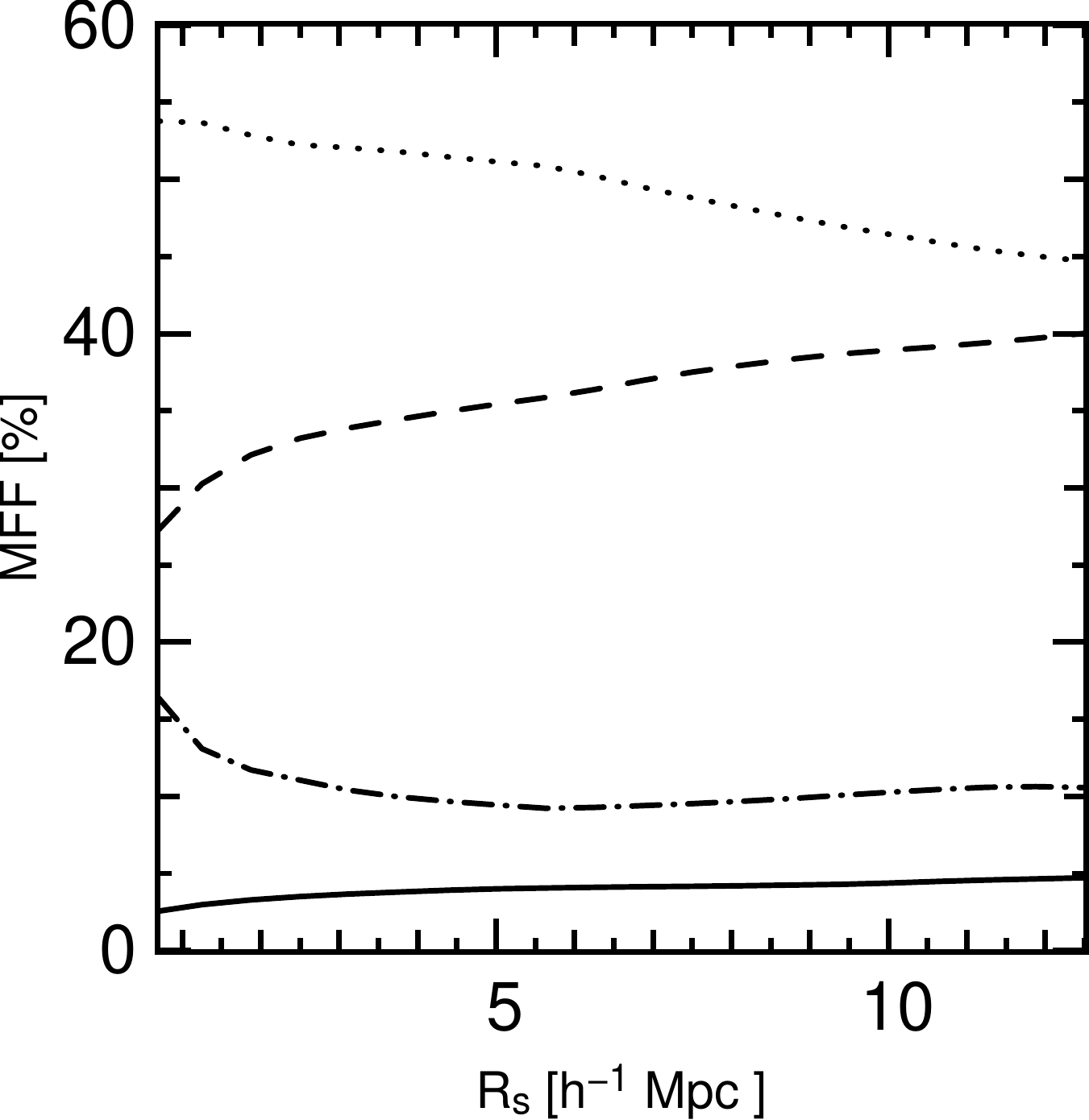}\hspace{0.5cm} 
\includegraphics[width=7cm]{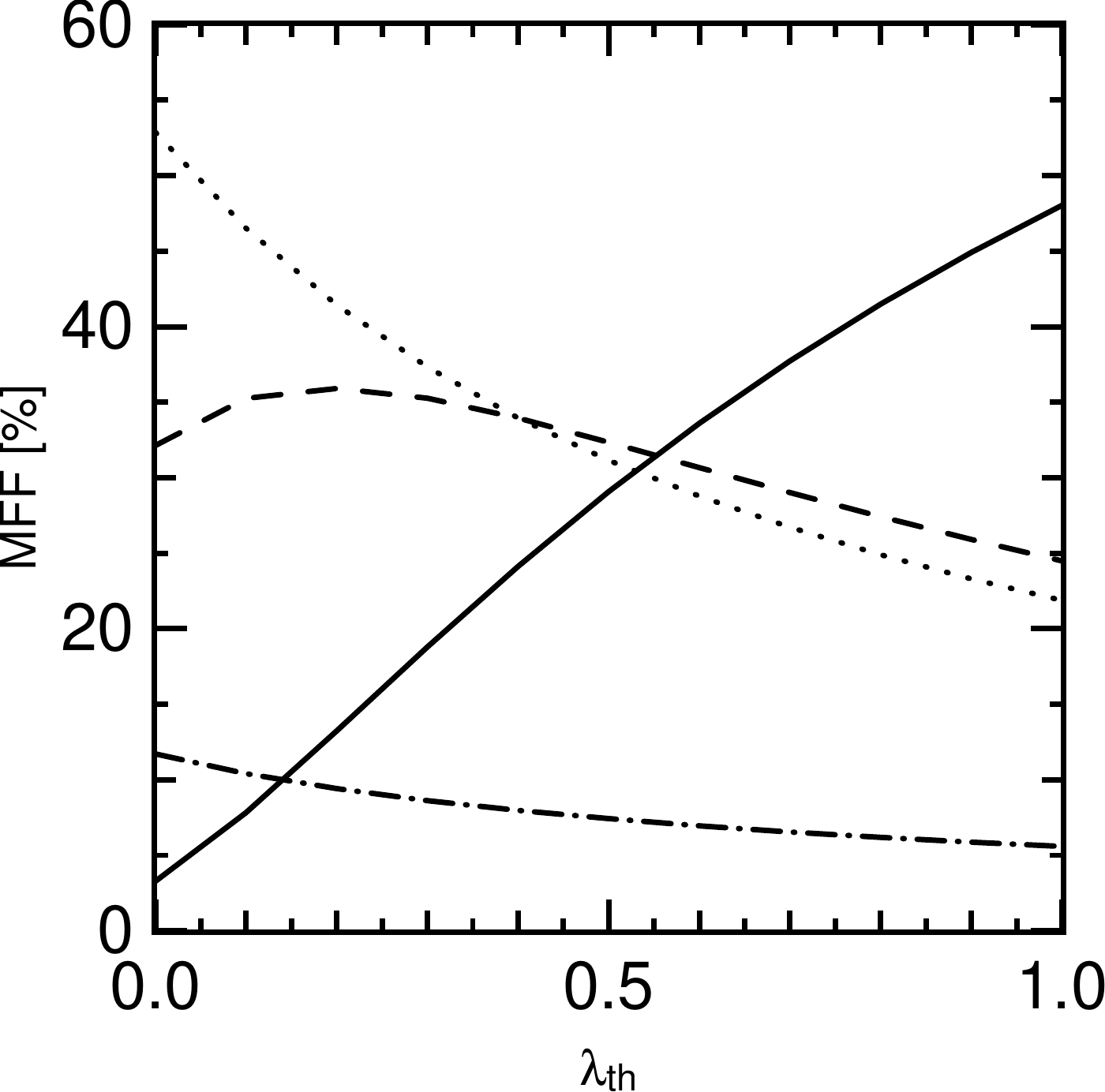} 
\caption{\emph{Upper panel}: the volume filling fraction of peaks, filaments, sheets and voids as a 
  function of the smoothing scale $R_\mathrm{s}$ for $\lambda_\mathrm{th}=0.0$ 
  (left), and as a function of $\lambda_\mathrm{th}$ for $R_\mathrm{s}=1.95 \hMpc$ 
  (right).  
  Continuous line: voids, dashed: sheets, dotted: filaments, 
  dotted-dashed: knots. \emph{Lower panel}: same as the 
  upper panel but for the mass filling fraction.}   
\label{fig:factors} 
\end{center}
\end{figure*}

\begin{figure*} 
\begin{center} 
\includegraphics[width=5cm]{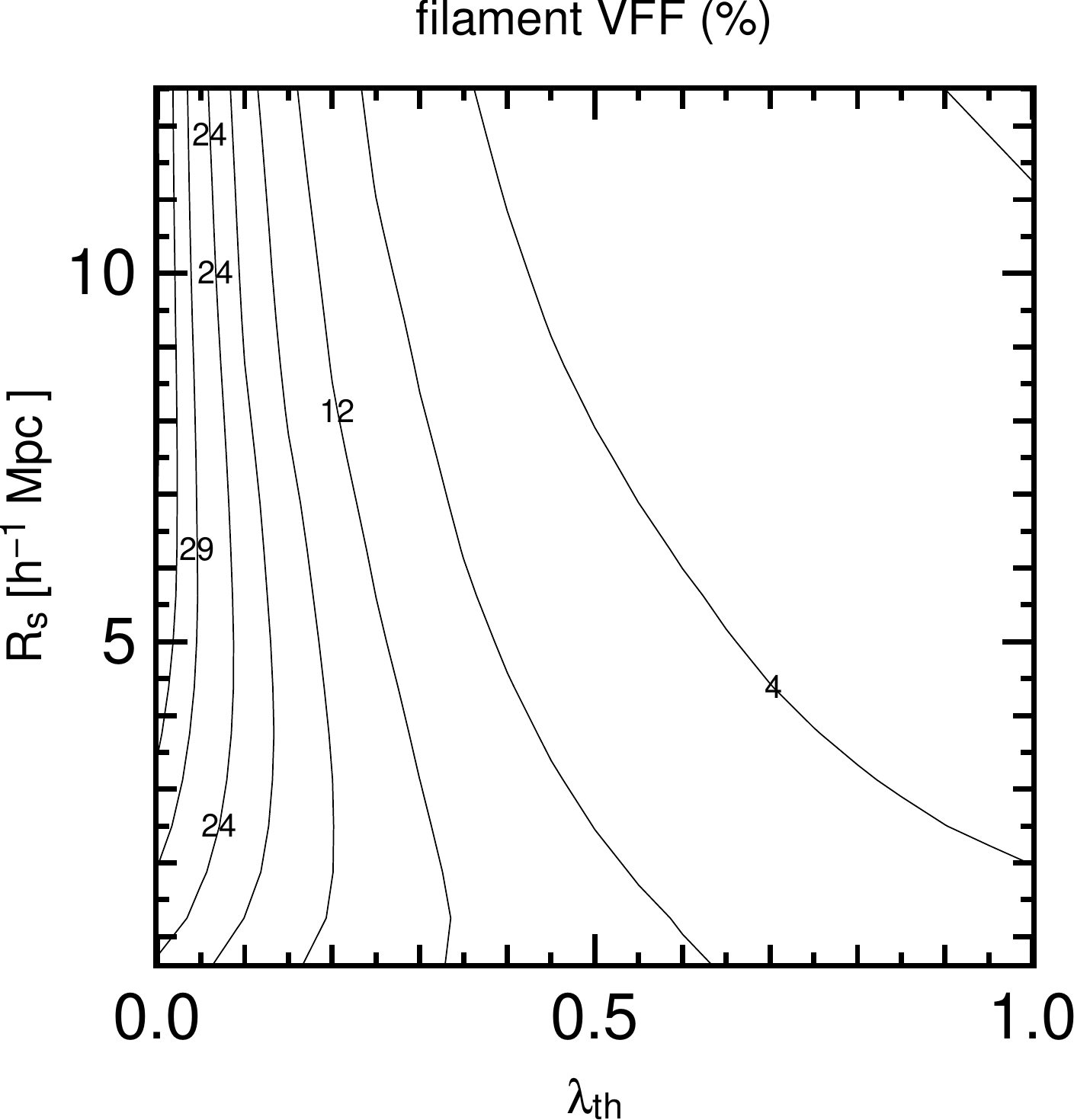}\hspace{1cm} 
\includegraphics[width=5cm]{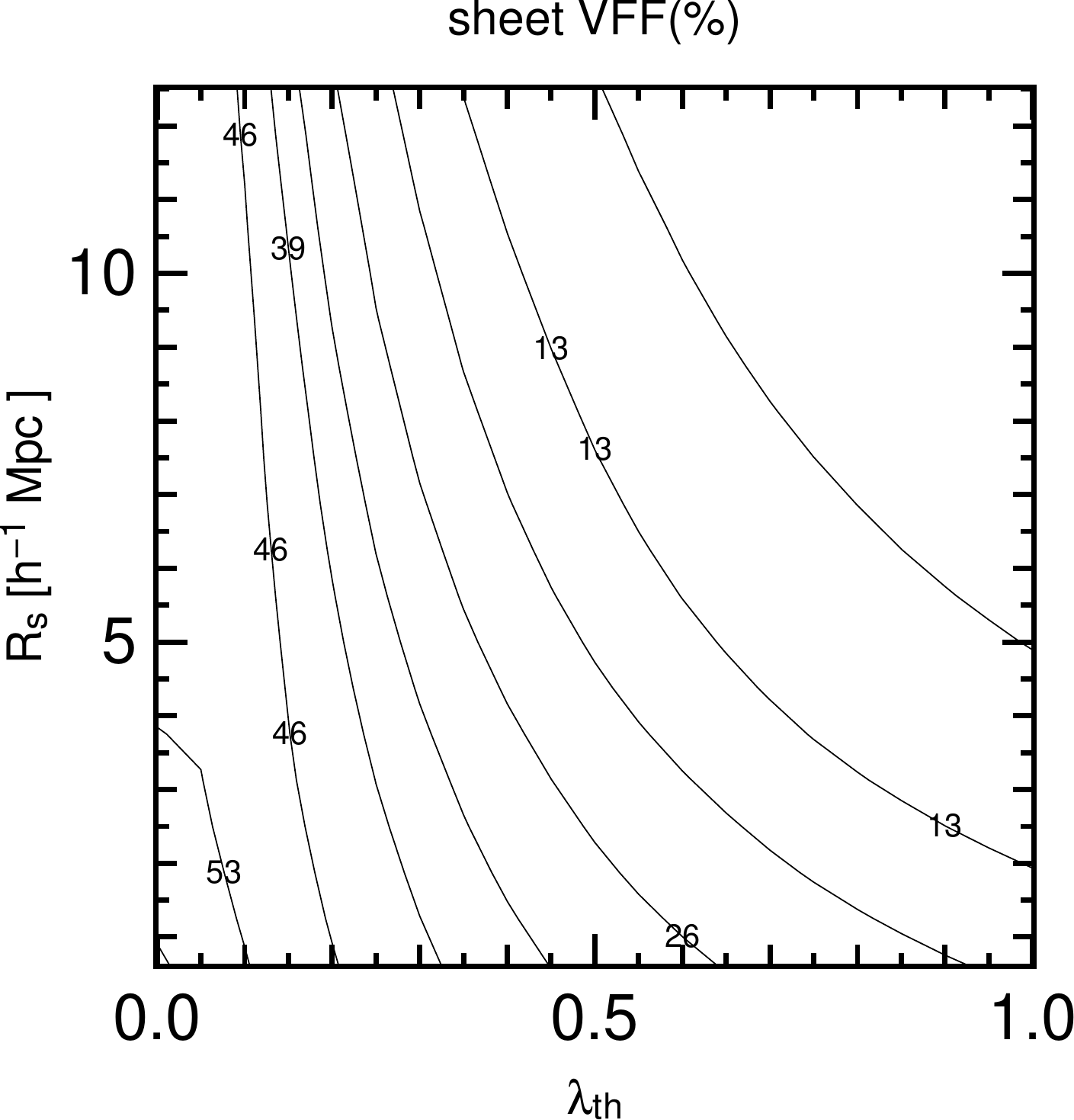}\hspace{1cm} 
\includegraphics[width=5cm]{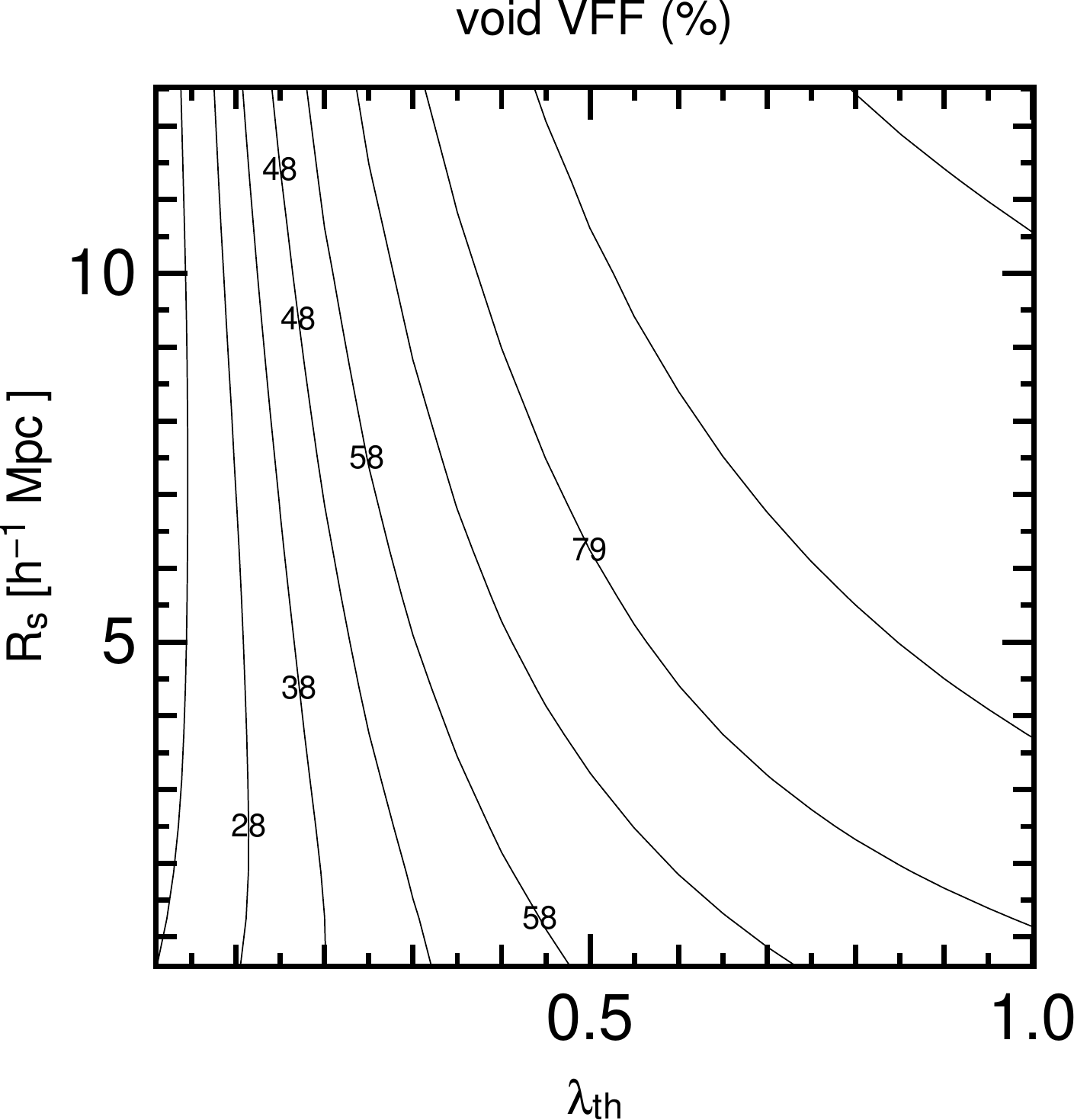}\\ 
\vspace{0.5cm} 
\includegraphics[width=5cm]{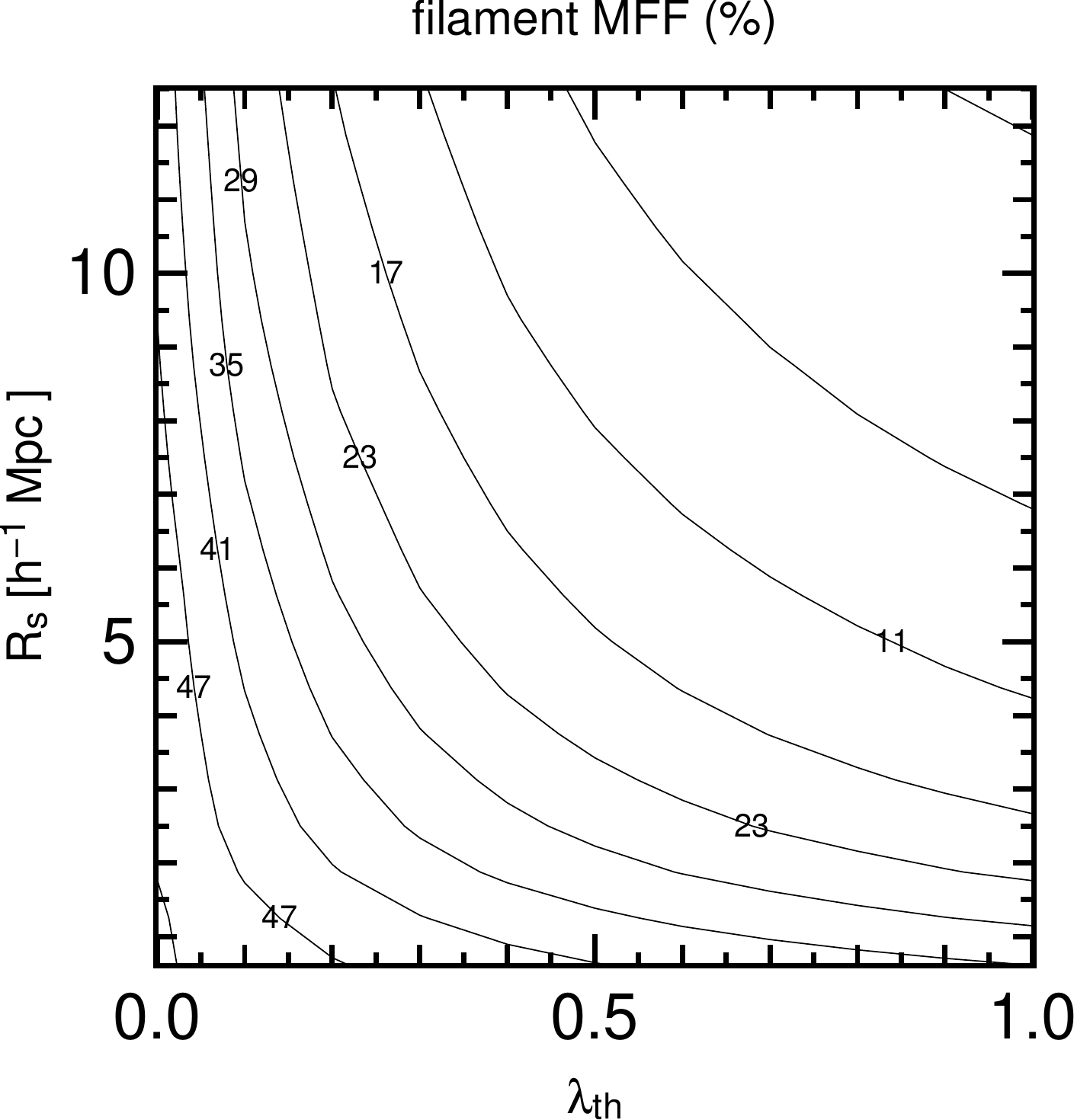}\hspace{1cm} 
\includegraphics[width=5cm]{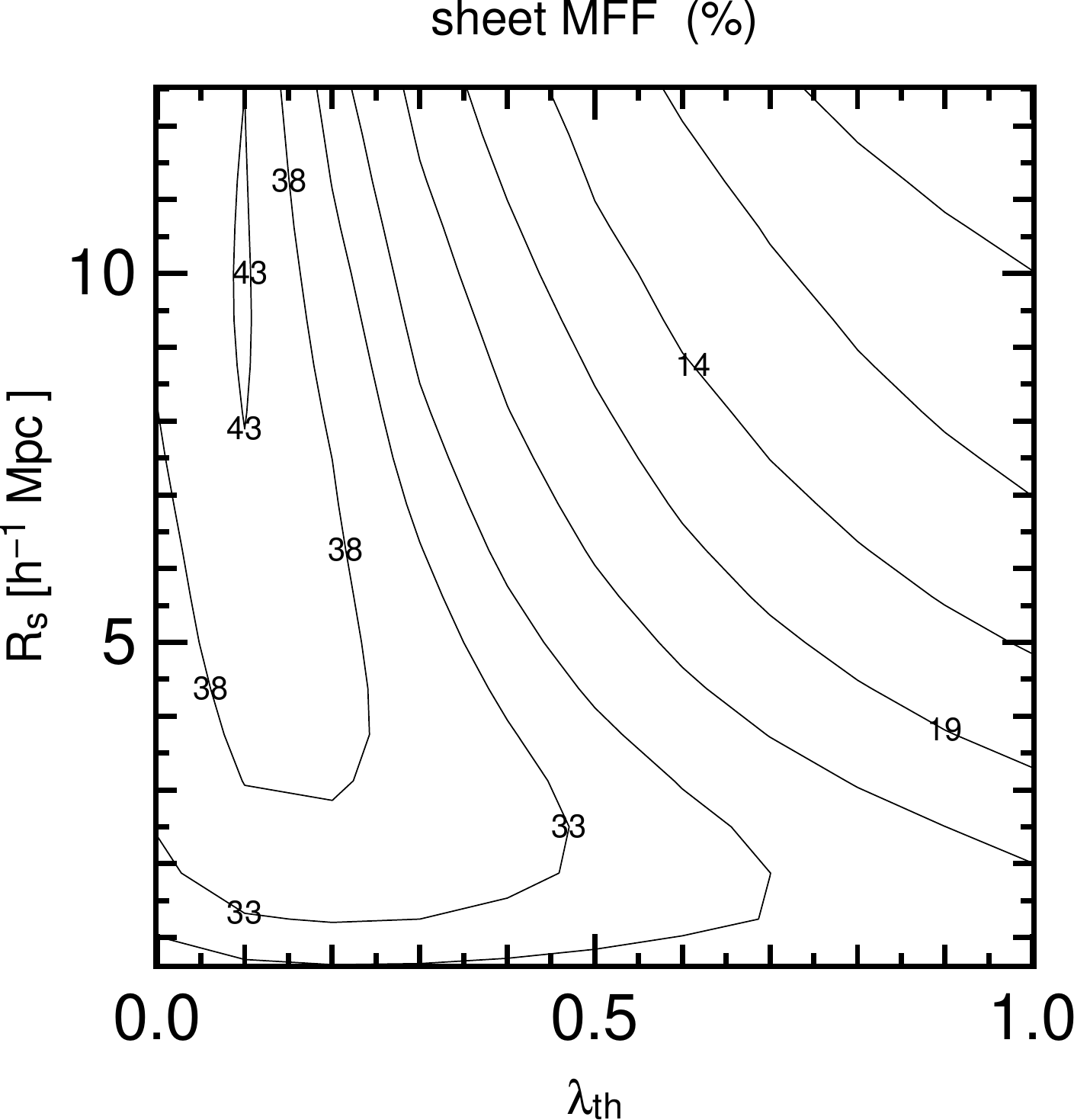}\hspace{1cm} 
\includegraphics[width=5cm]{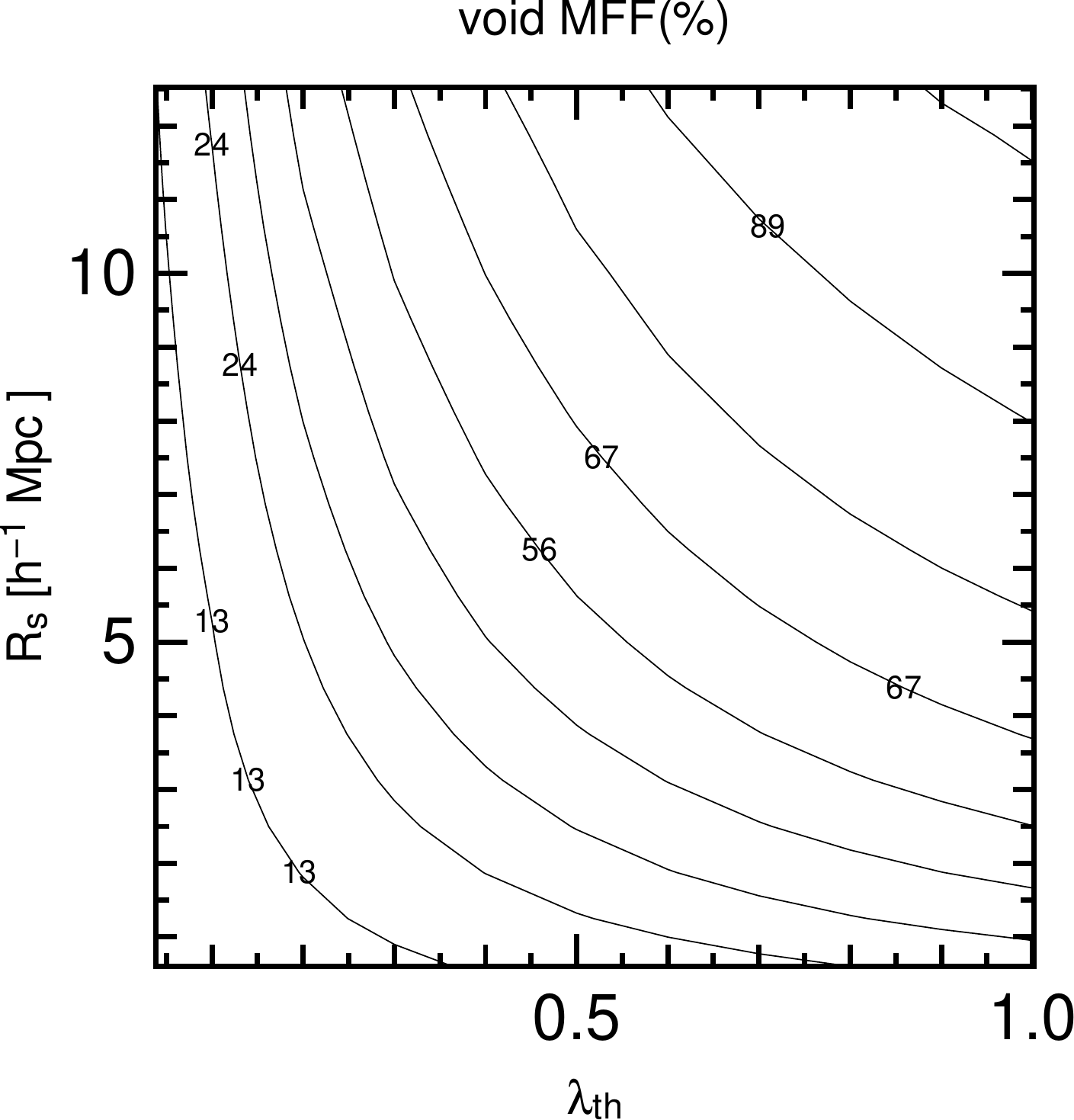} 
\caption{\emph{Upper row}. Isocontours for the volume filling fractions for the different kinds of 
  environment, in the plane $R_\mathrm{s}$-$\lambda_\mathrm{th}$.  
\emph{Lower row}. Isocontours for the mass filling fractions for the different kinds of 
  environment, in the plane $R_\mathrm{s}$-$\lambda_\mathrm{th}$.  In 
  both rows, from left to right: filaments,  sheets and voids. The 
  label on each isocontour is the percentage of the   total volume 
  filled by a given kind of environment.  
} 
\label{fig:planes} 
\end{center} 
\end{figure*}

The web classification depends on two  parameters that determine the 
environment. The first is the smoothing scale $R_\mathrm{s}$, and the second is the 
threshold for the eigenvalues $\lambda_\mathrm{th}$.  The dependence of the 
volume and mass filling fractions (VFF and MFF, respectively)  on these two 
parameters  is studied here. This is done for the four web types.

The VFF and MFF  are measured for every kind of environment.  First by fixing 
$\lambda_\mathrm{th}=0$ and varying the smoothing scale $R_\mathrm{s}$ between 
$0.625 \hMpc$ and $12.4 \hMpc$. Later, fixing the smoothing scale to $1.95 \hMpc$
and varying  $\lambda_\mathrm{th}$ between $0$ and $1$ with steps of
$0.1$. The results of these two kinds of cuts are shown in Figure \ref{fig:factors}.   
The VFF and MFF of the case $\lambda_\mathrm{th}=0.0$ and $1.0$, with 
two different smoothing scales, $R_\mathrm{s} = 0.625 \hMpc$ and $R_\mathrm{s} =
1.95 \hMpc$ are presented in Table \ref{tab:factors}. The  evolution of the
filling fraction with $\lambda_\mathrm{th} = 0$ reproduces the  asymptotic
results expected for large smoothing scales, which is $0.42$ for  sheets and
filaments and $0.08$ for voids and knots \citep{1970Ap......6..320D}.

\begin{table*} 
  \begin{center} 
    \begin{tabular}{c cccc cccc} 
      \hline\hline 
      &\multicolumn{4}{c}{$R_\mathrm{s}=0.625\hMpc$} &\multicolumn{4}{c}{$R_\mathrm{s}=1.95\hMpc$} \\  
      \hline
      &\multicolumn{2}{c}{$\lambda_\mathrm{th}=0.0$} 
      &\multicolumn{2}{c}{$\lambda_\mathrm{th}=1.0$}
      &\multicolumn{2}{c}{$\lambda_\mathrm{th}=0.0$}
      &\multicolumn{2}{c}{$\lambda_\mathrm{th}=1.0$} \\

      \hline 
      Web type  & Volume & Mass & Volume & Mass & Volume & Mass & Volume & Mass \\
      \hline 
      void        & 0.16 &     0.02 &  0.76 &  0.28  & 0.13 & 0.03 & 0.82 & 0.47\\  
      sheet      & 0.60 & 0.27 &  0.18 &  0.25  & 0.56 & 0.32 & 0.14 & 0.25\\ 
      filament  & 0.24 &    0.54 &  0.05 &  0.35 & 0.28 & 0.52 & 0.04 & 0.22\\ 
      knot        & 0.01 &  0.16  &  5.0e-3   & 0.12  & 0.01 & 0.11 & 2.8e-3 & 0.06\\ 
      \hline 
    \end{tabular} 
    \caption{Volume and mass filling fractions for the four web types for two
      different smoothing scales, $R_\mathrm{s}=0.625\hMpc$ and
      $R_\mathrm{s}=1.95\hMpc$. For each smoothing two extreme values of the
      threshold are used, $\lambda_\mathrm{th}=0.0$ and
      $\lambda_\mathrm{th}=1.0$. The volume filling fractions are similar for the same
      values of the threshold $\lambda_\mathrm{th}$ regardless of which
      smoothing scale is used.}   
    \label{tab:factors} 
  \end{center} 
\end{table*}

The most striking feature that emerges from  Figure \ref{fig:factors}  is that 
of the strong dependence of VFF and MFF  of the voids on 
$\lambda_\mathrm{th}$. This is to be contrasted with the other web types which 
show quite a similar behavior with the change of the smoothing and the change 
of the threshold. It follows that the voids can serve as a sensitive monitor 
and indicator of the cosmic web. For the case of a null threshold the 
dependence on the smoothing length is weak. The increase of $R_\mathrm{s}$ 
corresponds to a transition to the linear regime where the density field is 
closer to be Gaussian. 
 
Then we perform an exploration of the space
$\lambda_\mathrm{th}$-$R_\mathrm{s}$ by making the environment classification with 
steps of $\Delta\lambda_\mathrm{th}=0.1$ for each  smoothed density field. We 
measure the filling fractions for each couple 
$R_\mathrm{s}$-$\lambda_\mathrm{th}$.  The results are shown in Figure 
\ref{fig:planes}. The important result here  is that in the range of explored 
values, the threshold  on the eigenvalues is more important that the smoothing 
scale to fix the volume filling fraction.

Inspection of the behavior of the VFF and MFF in the  $R_\mathrm{s}$
- $\lambda_\mathrm{th}$ plane reveals that voids on the one hand and the sheets and 
filaments on the other hand are complementing one another. Meaning that the
VFF and MFF for voids increase  with both  $R_\mathrm{s} $ and $
\lambda_\mathrm{th}$ at expenses of the  volume and mass in sheets and filaments.

This provides another evidence for the distinct nature of 
voids. The VFF of the voids, sheets and filaments clearly show the 
significance of the $\lambda_\mathrm{th}  \lesssim 0.2$ case.  For that 
threshold the contours lines are almost vertical, implying that the 
VFF of these web elements are almost independent of $R_\mathrm{s}$.    
The MFF, on the other hand, distinguishes very easily different values of the
smoothing scale $R_\mathrm{s}$.

\section{Percolation of Voids} 
\label{sec:res-void}

\begin{figure*} 
\begin{center} 
\includegraphics[width=7cm]{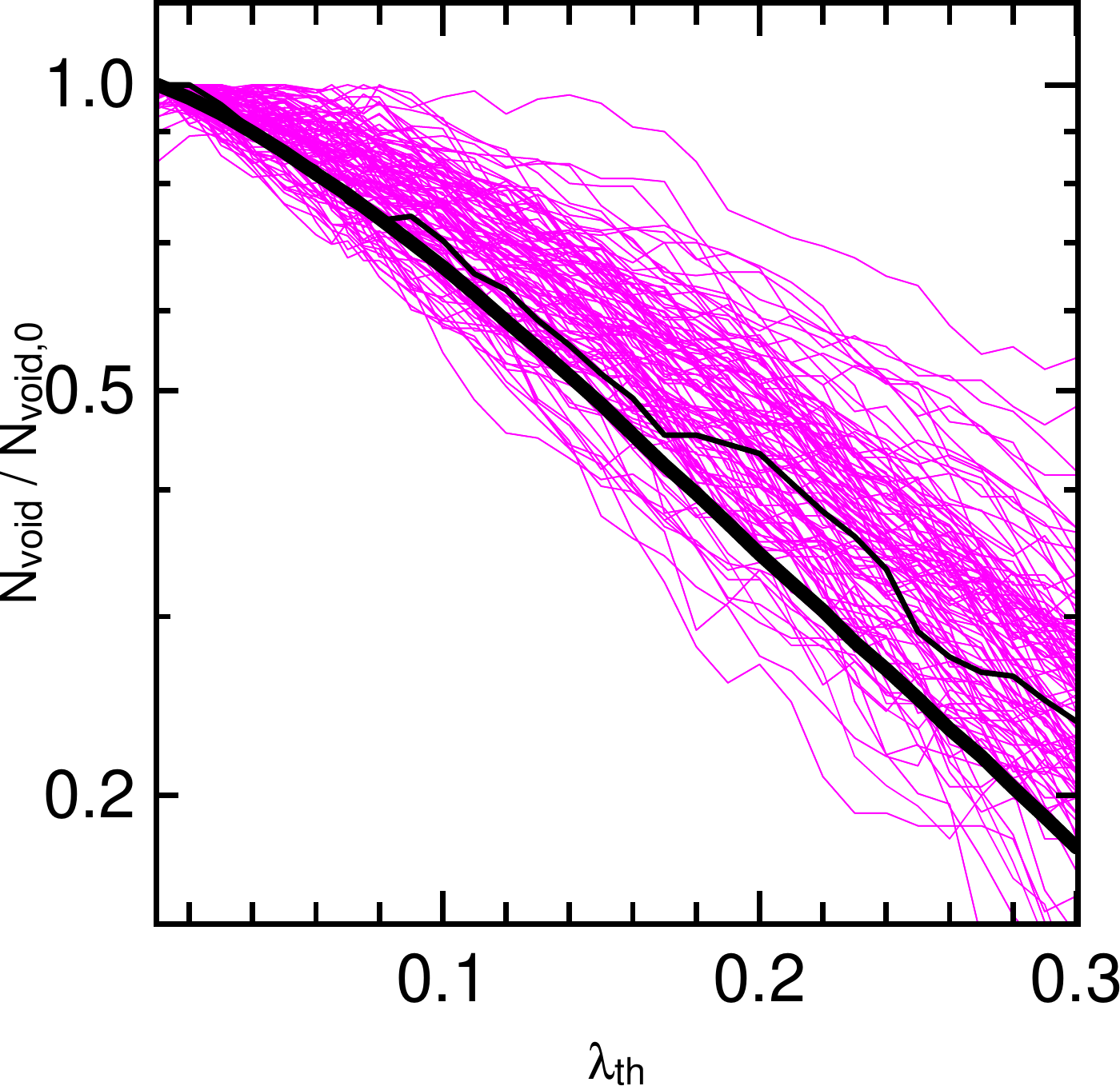}\hspace{0.5cm} 
\includegraphics[width=7cm]{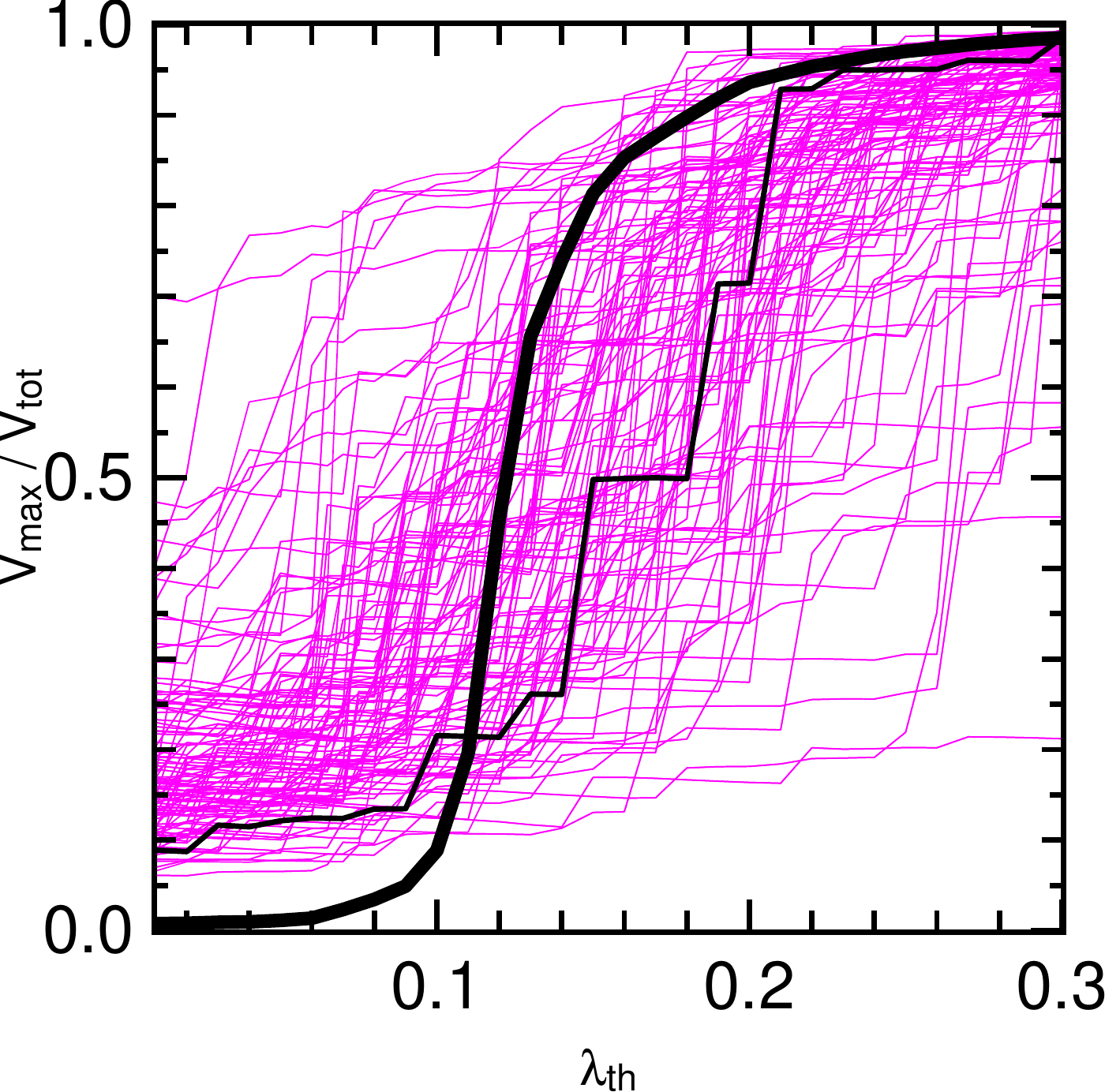} 
\caption{Percolation results. \emph{Left panel}: number of voids as a function of the threshold in
  eigenvalues normalized to the number of
  voids at $\lambda_\mathrm{th}=0.0$. 
  A fixed smoothing of $R_\mathrm{s }= 1.95 \hMpc$ is used here for all the simulations. 
  The thick line shows the results
  for  the $1 \hGpc$ simulation, the thin line shows the  results for the $160 \hMpc$
  simulation. The gray lines show the results for the   sub-volumes extracted from
  the $1 \hGpc$ simulation.   \emph{Right panel}: fraction of the total void volume
  occupied by the most voluminous void. The line coding is the same as in the
  left panel. Results from the large $1 \hGpc$ differ greatly from the results
  in  the simulation $160 \hMpc$, nevertheless the later is consistent within the scatter
 deduced from the sub-volumes. Even when it is clear that the detailed shape
 of the curve depends on the simulation size, it seems to be a robust feature that
 the largest change in the super-void size is presented around
 $0.1 \lesssim \lambda_\mathrm{th} \lesssim 0.2$.}
\label{fig:fracciones} 
\end{center} 
\end{figure*}

Given that each grid node is flagged  as belonging to a void, sheet, filament or a knot one 
can define new objects made by connecting grid nodes of the same type. In 
particular we proceed here to define a void as an object made of the 
collection of neighboring void type grid nodes. Neighboring nodes are 
associated  by the FoF  algorithm to form individual objects  
as described at the end of \S \ref{sec:n-body}. 

Here we focus on 
the analysis of the statistical distribution of the sizes of the voids and 
their percolation.  The emphasize on voids does not stem only from the 
extensive work done on their  properties   \citep{2008MNRAS.387..933C} but 
also because  they constitute the most  sensitive  gauge of the cosmic web and its 
dependence on the threshold of the eigenvalues. Filaments have
received attention as well in the literature
\citep{2006MNRAS.366.1201N}, in the next section we present a brief analysis
of their properties.

A simplified characterization was already performed in \S \ref{sec:res-web} 
by measuring the volume and mass occupied in the void environment.  Here, the 
dependence of the number of voids and   their percolation properties on 
$\lambda_\mathrm{th}$ is examined. The percolation is quantified now by  the 
fraction of volume of the most voluminous void to the total volume 
occupied by all voids.  
 
 The void identification is performed here at a fixed  smoothing scale 
$R_\mathrm{s}= 1.95 \hMpc$ and by varying the eigenvalues threshold in the 
range of  $0<\lambda_\mathrm{th}<0.3$.   In the Figure \ref{fig:fracciones}
(left panel) one can see that the number of voids in the simulation roughly decays
exponentially as a function of the threshold parameter $\lambda_\mathrm{th}$,
$N_{void}  = N_{0} \exp(- \lambda_\mathrm{th}/\lambda_{D})$, where $N_{0}$ is
the number of voids for the null threshold and  $\lambda_{D}$.

Figure \ref{fig:fracciones} (right panel) presents the fraction of the 
volume of the largest (in volume) void to the total volume of all the
voids. A transition occurs between $0.1 \lesssim \lambda_\mathrm{th} \lesssim
0.2 $  where it jumps from a ratio $\leq  0.1$ to $\geq  0.9$. Note that
the percolation starts at the stage in which the  VFF of voids is only $\sim
25\%$.  It follows that  in spite of  the  small VFF obtained for $\lambda_\mathrm{th} \sim 0.1$ the 
voids start to coalesce and form one super-void which encompasses $90\%$ of 
the total volume of voids when the void VFF reaches $60\%$. As we will show in
\S \ref{sec:cosmic-variance}, this transitional scale is dependent on the
particular simulation under consideration.

\section{Fragmentation of Filaments} 
\label{sec:frag-filament} 

\begin{figure} 
\begin{center}
\includegraphics[width=7cm]{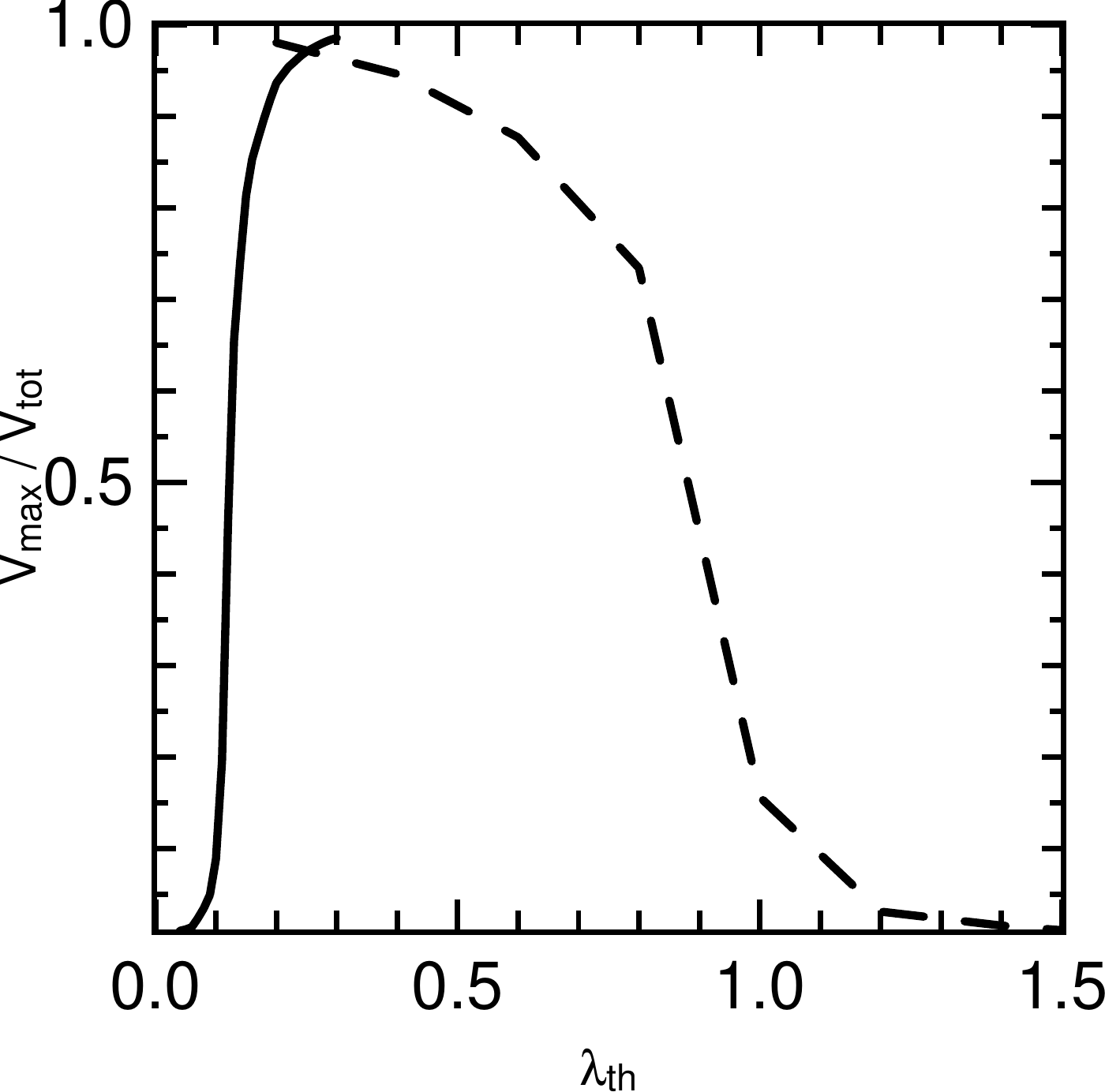}\hspace{0.5cm} 
\caption{The continuous line shows the fraction of total volume in voids occupied by the largest
void. The dashed line shows the fraction of the total volume in filaments occupied by the largest
filament. Both curves refer to the  $1 \hGpc$ simulation. As
the eigenvalue threshold rises the voids percolate, while the network of
filaments fragments. The two fractions are the same at
$\lambda_\mathrm{th}=0.25$. } 
\label{fig:perc_comparison}
\end{center}
\end{figure}

Along with voids, filaments have received some attention in
the literature as a natural way to probe large scale structure
\citep{2008MNRAS.383.1655S}. A study of the filaments is performed here,
focusing on their percolation dynamics.  

In the opposite sense to voids, filaments start to fragment as the threshold
$\lambda_\mathrm{th}$ is raised.  This is clear in  Figure
\ref{fig:dens_field}, which shows that the filaments dissappear as
$\lambda_\mathrm{th}$  increases, making room for the growing voids.  

Consequently, one could expect that for some range of $\lambda_\mathrm{th}$
there are two coexisting environments: a complex of percolating voids and
a network of interconnected filaments, something close to the visual
impression of the comic web. The percolation analysis of the filament
and void networks can help to define an interval of $\lambda_\mathrm{th}$
values where the environment studies would be feasible.

We use the $1\hGpc$ simulation following the approach of the previous section, Figure
\ref{fig:perc_comparison} shows the
ratio of the volume of largest filament to the total volume occupied
by filaments, overplotted is the same fraction for voids (presented in
the right panel of Figure \ref{fig:fracciones}). The filament fraction evolves
from a value close to $\sim 1$ for $\lambda_\mathrm{th}=0.0$ down to values close to zero for
$\lambda_\mathrm{th}\sim 1.5$. It confirms the visual intuition we had of a fully 
interconnected network of filaments that fragments as the eigenvalue threshold
increases.

The percolation/fragmentation curves intersect at $\lambda_\mathrm{th}=0.25$,
when the two fractional volumes are $\sim 97\%$.  Heuristically, one can
assume a given network to exist, namely percolate, when its fractional volume
exceeds $95\%$.  The web is then defined at the threshold level at which the
voids and filaments networks coexist. This implies a threshold interval of
$0.20\lesssim\lambda_\mathrm{th}\lesssim 0.40$.  Such a heuristic approach is in a good
agreement and matches the visual impression of the LSS (Figure
\ref{fig:dens_field}).  This approximate interval
should hold for lower smoothing scales, as the volume filling fraction
(our gauge for the percolation dynamics) is almost independent of the
smoothing scale for the range of thresholds considered, as seen in Figure
\ref{fig:planes}, and suggested as well in the values of Table \ref{tab:factors}.

\section{Simulation Box Size and Cosmic Variance}
\label{sec:cosmic-variance} 
 
\begin{figure*} 
\begin{center}
\includegraphics[width=7cm]{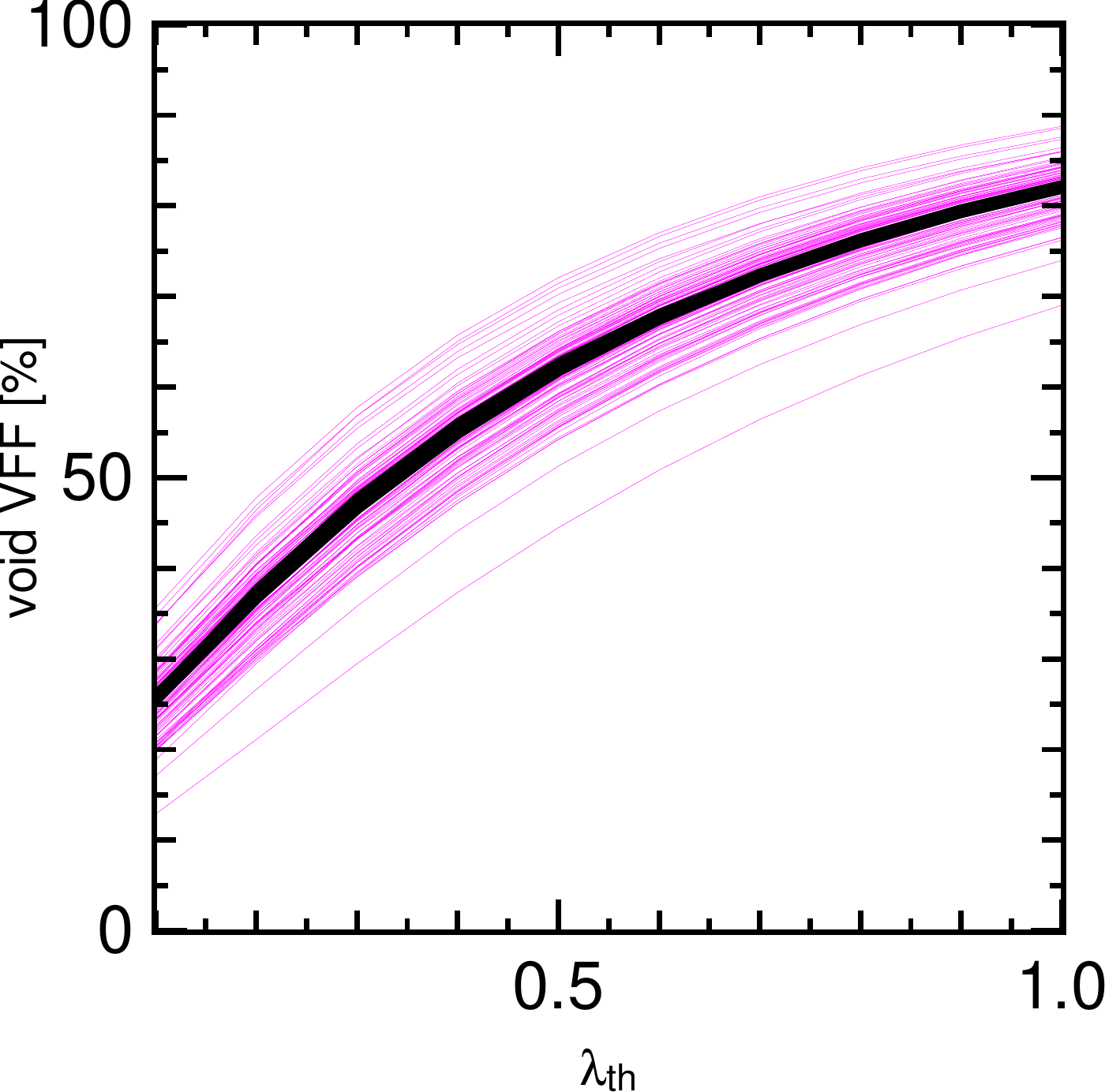}\hspace{0.5cm}
\includegraphics[width=7cm]{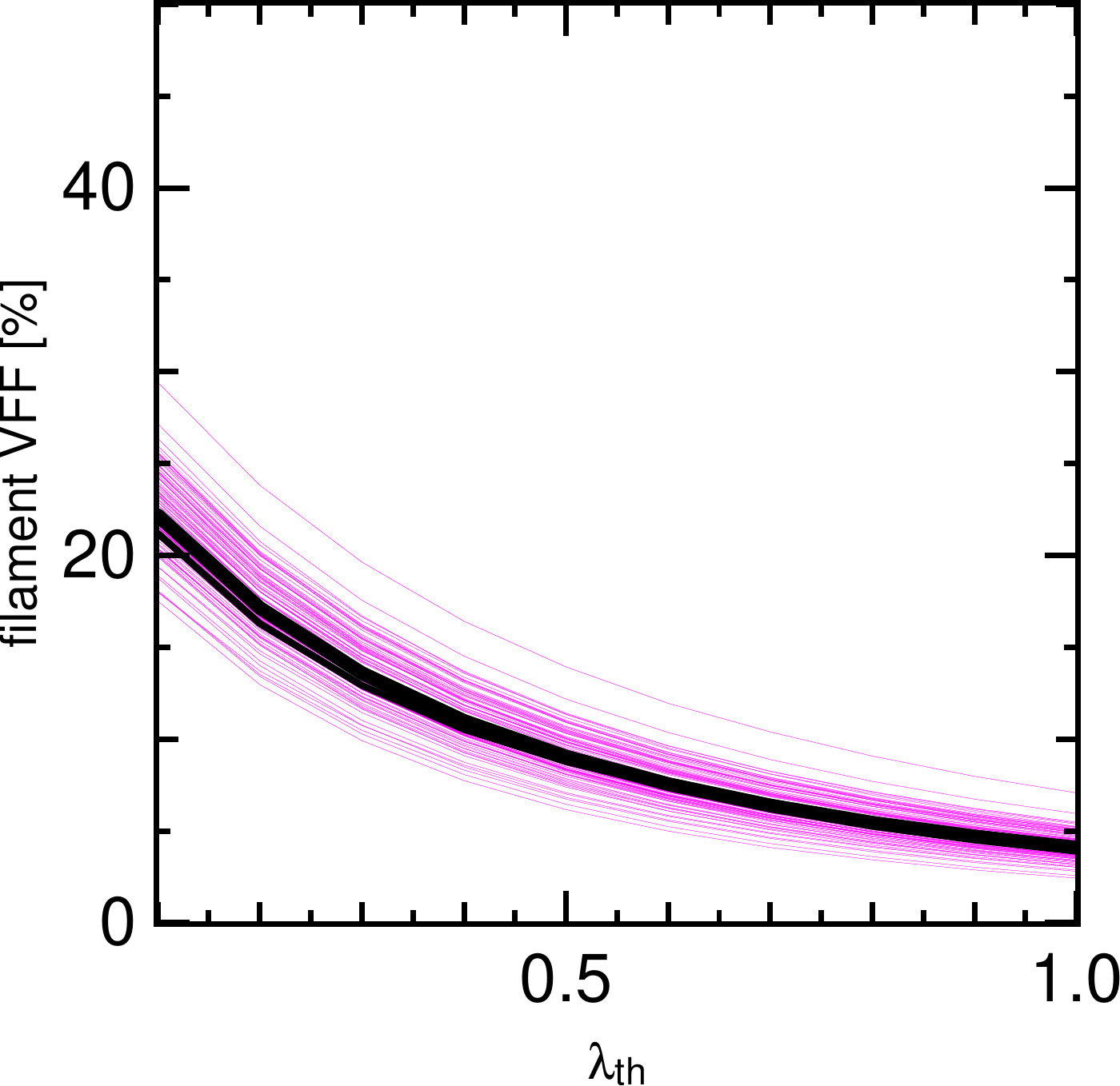}\\
\vspace{0.5cm}
\includegraphics[width=7cm]{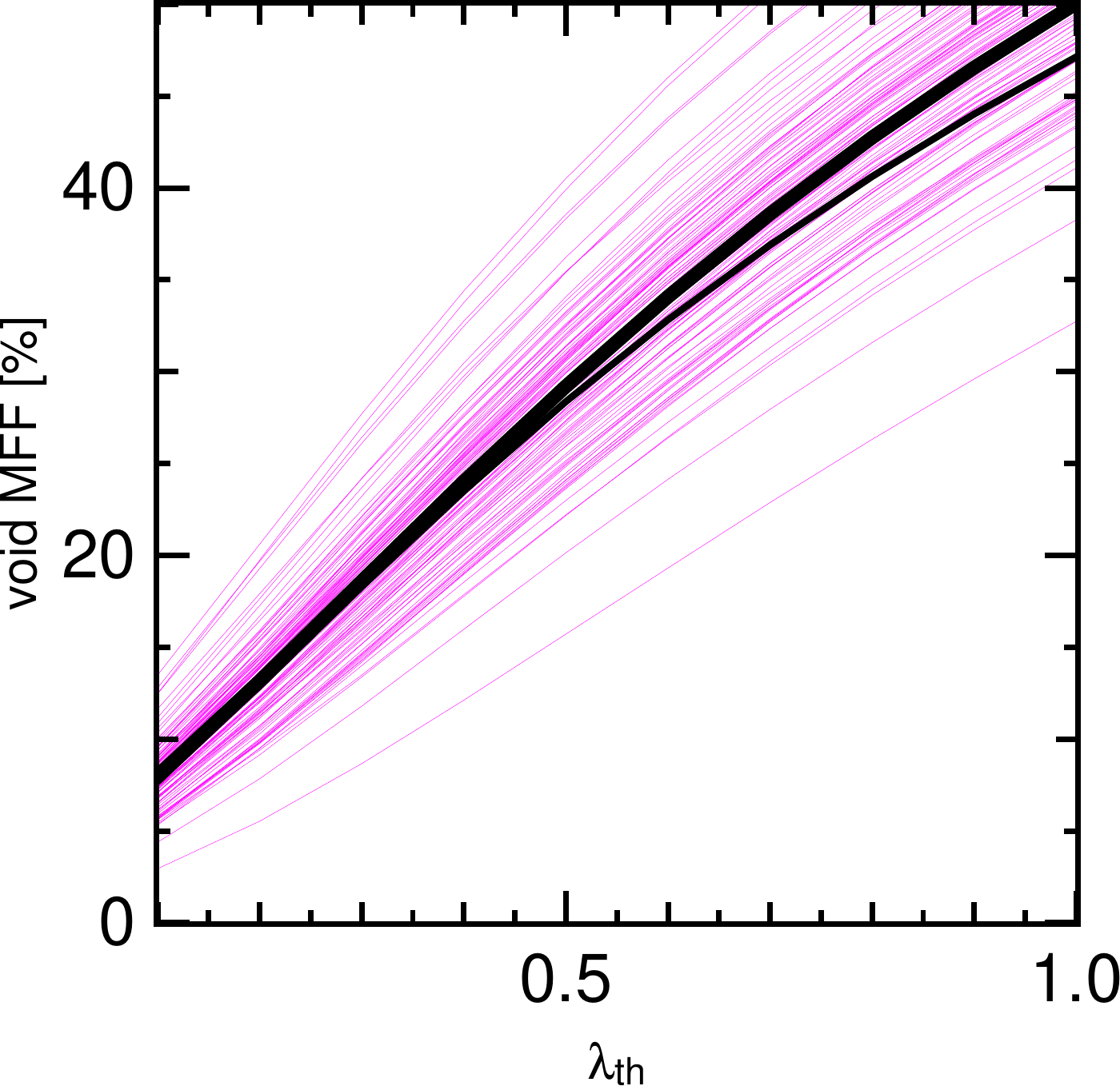} \hspace{0.5cm}
\includegraphics[width=7cm]{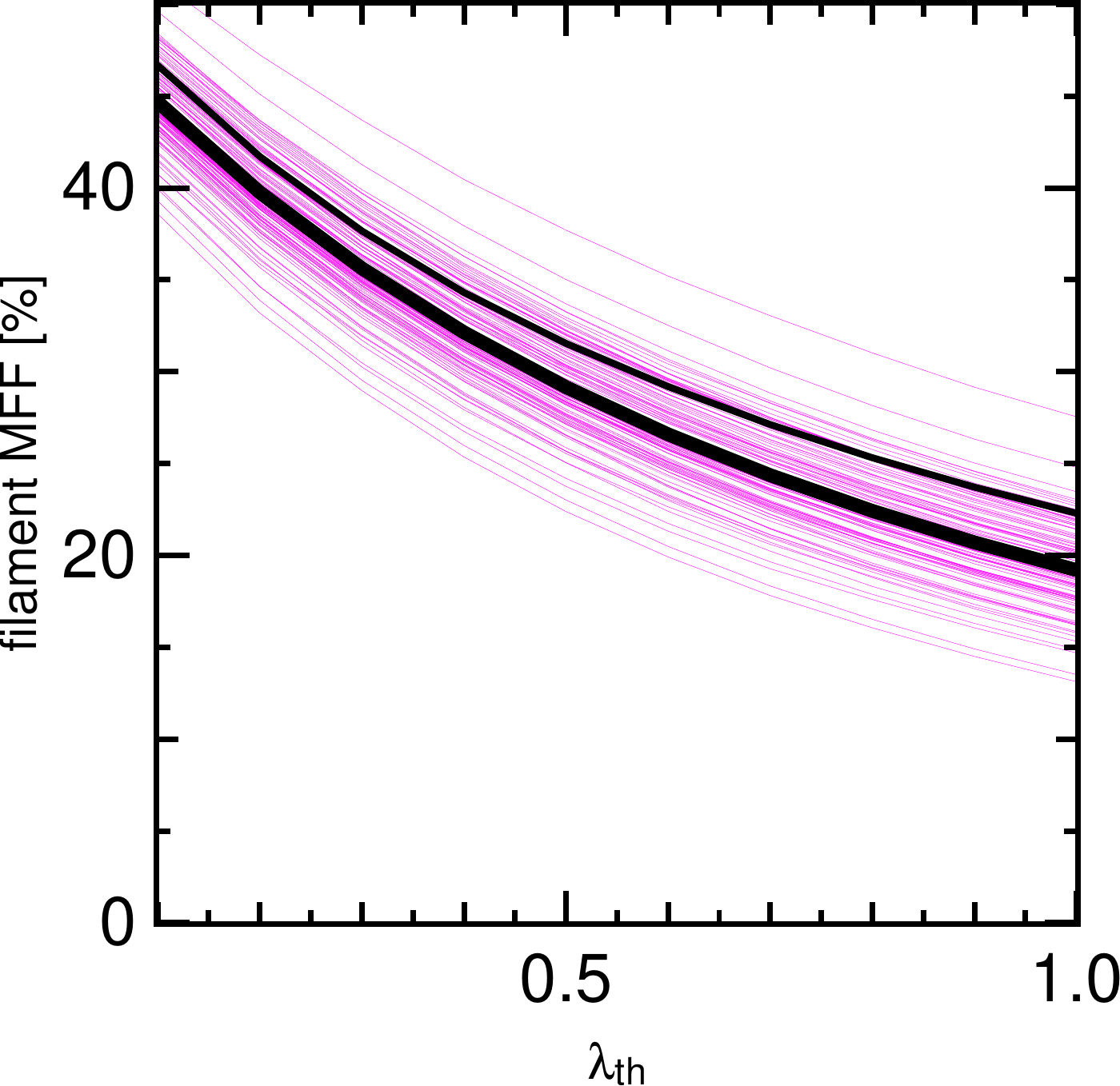} 
\caption{Cosmic variance effects on volume and mass filling
  fractions for voids and filaments. Thick black line: simulation $1 \hGpc$, thin
  black line: simulation $160 \hMpc$, gray lines: sub-volumes extracted from the $1
  \hGpc$ simulation. \emph{Upper panels}: volume filling fraction of voids (left) and
  filaments (right) as a function of the
  eigenvalue threshold $\lambda_\mathrm{th}$. \emph{Lower panels}: same as the
  upper panels for the mass filling fractions.}
\label{fig:FF_voids} 
\end{center} 
\end{figure*}

In order to quantify the influence of the simulation box size and the
impact of cosmic variance we measure for three different kind of simulations the MFF and VFF for voids and filaments,
and the percolation of voids. The first is the $1\hGpc$ simulation with the density field interpolated over a $512^3$ grid
and smoothed with $R_\mathrm{s}=1.95\ \hMpc$, used to quantify the effect of
simulation size. The second is the $160
\hMpc$ simulation interpolated over a $256^3$ grid smoothed over the 
same physical scale. The third kind is the $1 \hGpc$ simulation split in
smaller non-intersecting cubes of $160 \hMpc$, used to quantify the
effect of cosmic variance.

Figure \ref{fig:FF_voids} shows the result for the VFF and MFF. The results for
the $160 \hMpc$ simulation are well within the variance calculated from the
sub-volumes in the $1 \hGpc$ simulation. In the case of the MFF,
the result for the $160\hMpc$ are located far from the mean value in the $1
\hGpc$,  nevertheless it is located with in the dispersion defined
by the sub-volumes. In general, the VFF and MFF are consistent in all the
three kinds of simulations. The agreement is less impressive than with the
  VFF, perhaps due as well to the different values of $\sigma_8$ used in the
  simulations, the large simulation has $\sigma_8=0.79$, while the small simulation has
  $\sigma_8=0.75$

The growth of the largest void in the simulation, Figure \ref{fig:fracciones},
is very different between the three kinds of simulations. The fraction
of void volume occupied by the super-void 
is very dependent on the simulation size. For the large $1 \hGpc$ simulation, the initial
values of $V_{max}/V_{tot}$ for $\lambda_\mathrm{th} = 0.0$ are the
lowest possible, this can be readily understood by the fact that $V_{tot}$
grows with the simulation box size, while $V_{max}$ should be on the same
order of magnitude regardless of the simulation size. For values larger than
$\lambda_\mathrm{th}=0.2$ the large simulation has almost percolated into a
single super-void, while the sub-volumes still show a large dispersion in their
percolation behavior. The results for the
small $160 \hMpc$ simulation are consistent with such dispersion. In spite of
that, in all the three cases there is a clear transitional behavior starting
at $\lambda_\mathrm{th} = 0.1$ and finishing around $\lambda_\mathrm{th} =
0.2$, even when the detailed behavior with $\lambda_\mathrm{th} $ is far from
being the same.

\section{Concluding Remarks} 
\label{sec:disc}

This paper presents an improved method to identify large scale environment in dark 
matter simulations. Our scheme is based on the analysis of the Hessian of 
the gravitational potential generated by the dark matter distribution. 
The algorithm presented here constitutes an improvement on 
the scheme of  \cite{2007MNRAS.375..489H}, involving a pertinent
reinterpretation of the dynamics in the problem.

The change is done through the addition of a free parameter
$\lambda_\mathrm{th}$ related to the dynamical time associated with the
collapse. This important improvement allows a more realistic treatment of 
the cosmic web.

Inspection of the different plots of Figure \ref{fig:factors} reveals the 
striking difference in the way the cosmic web, and in particular the voids, 
respond to the changes in the Gaussian smoothing and the $\lambda_\mathrm{th}$'s 
threshold. Keeping a null $\lambda_\mathrm{th}$  and changing $R_\mathrm{s}$ 
we see that the non-linear evolution does not change the ranking of the VFF 
and MFF found in the deep linear regime (i.e., $R_\mathrm{s}\approx 
12.5 \hMpc$). Namely, for the null threshold the sheets have the highest VFF and the 
filaments the highest MFF, independent of $R_\mathrm{s}$. Considering the case 
of a fixed $R_\mathrm{s}=1.95\hMpc$, the void VFF 
grows strongly with $\lambda_\mathrm{th}$ and above   $\lambda_\mathrm{th} 
\sim 0.5$  the voids have the highest  VFF.  The results on the VFF are
extremely robust respect to the simulation size and cosmic variance effects. In the
case of void MFF, it changes  from the lowest one at  $\lambda_\mathrm{th}=0$ to the second highest  at 
$\lambda_\mathrm{th} =1.0$.

The nature of the web changes dramatically with the  
threshold and it becomes volume dominated by the voids as 
$\lambda_\mathrm{th}$ increases. The MFF shows a larger cosmic variance,
compared with the VFF. Also, it is  equally sensitive to changes in the
smoothing scale and the threshold value.

The web classification provides  a set of flagged points on a grid and  
the collection of neighboring grid points of a given environmental 
type, connected by a FoF algorithm,  forms objects we call voids, sheets, 
etc. The statistical properties of the system of voids and their dependence on 
$\lambda_\mathrm{th}$ have been explored here.  In particular the number of 
isolated voids and their sizes have been analyzed, finding that 
the number of isolated voids roughly decreases exponentially with 
$\lambda_\mathrm{th}$. In the $1 \hGpc$ simulation
the system of voids percolates between  $0.1 \lesssim \lambda_\mathrm{th} \lesssim 0.2$,
at which  the largest void jumps from having less than $10\%$ to $90\%$ of
the volume occupied by voids.   The percolation dynamics is also seen on
average in the different smaller simulations, with a wide spread on the way
the percolation is observed, but keeping the same threshold interval for the
transition.

The association of the eigenvectors of the deformation tensor with the collapse
time, and hence with the age of the universe, enables in principle a
theoretical determination of $\lambda_\mathrm{th}$. Using the spherical
collapse model, calculated within the WMAP3 cosmology, the threshold value is
$\lambda_\mathrm{th}=3.21$. However, at that high value the web looks very
fragmented, in particular the network of filaments.  The application of the
ellipsoidal collapse model \citep{2002MNRAS.329...61S} might provide a better
theoretical estimate.

Short of a ``first principle'' determination of $\lambda_\mathrm{th}$ we resort
to a heuristic approach. We look for the range of $\lambda_\mathrm{th}$ over which the percolated
networks of the voids and filaments coexist for a fixed smoothing scale
of $R_\mathrm{s}=1.95 \hMpc$ in the $1 \hGpc$ simulation. The voids and filaments behave in
an opposite way in terms of the dependence of the percolation on the threshold
value. At low $\lambda_\mathrm{th}$ the voids are isolated and the 
filaments percolated and at high $\lambda_\mathrm{th}$ the voids percolate and the
filaments are fragmented. Adopting a  $95\%$ in the fractional volume as
defining the percolation transition we find the web to be defined by a
threshold in the range of  $0.20\lesssim\lambda_\mathrm{th}\lesssim 0.40$. This  range stands
in good agreement with the visual impression obtained from the simulations.

The notion of the cosmic web is not new. The filamentary structure has been 
extensively studied, mostly within the context of the Zeldovich pancakes
\citep{1970A&A.....5...84Z}. The role of voids has also been heavily studied
and many algorithms for voids  finding have been suggested (see
\cite{2008MNRAS.387..933C}). We have been motivated  by the computational
simplicity and the elegance of the \cite{2007MNRAS.375..489H}  approach and
have modified it in a way that reproduces the web as it emerges  from
observations and simulations. Our main drive is to provide a simple, fast  and
precise tool for classifying the environmental properties of each point in
space. Using  the non-zero thresholding of the eigenvalues of the Hessian   of
the potential indeed provides a very efficient tool that can be easily applied 
to simulations. 

The same analysis can be performed at different redshifts in the simulation,
allowing the classification of environment as a function of time. 

The dynamical nature of the web classification implies that the web type might affects the 
dynamical evolution of DM halos and of galaxy formation. This might  have profound consequences   on the star formation timescale....
\citep{2008arXiv0805.2191G}.  It follows that the web  classification can be
introduced into semi-analytical modeling, as a dynamical  tag 
that together with the mass of the DM halos dictate the the mode gas accretion onto galaxies.

A major challenge that is still to be addressed is the application of the
method to the distribution of galaxies. Short of that, the algorithm remains in
the theoretical realm of simulations and semi-analytical modeling of galaxy
formation.

\section*{Acknowledgments}

The support of the European Science Foundation through the ASTROSIM Exchange
Visits Program and of DAAD through the PPP program is acknowledged. 

The simulations were performed on the Leibniz Rechenzentrum Munich (LRZ),
partly using German Grid infrastructure provided by AstroGrid-D. 

We would like to thank  the DEISA consortium for granting us the computing time in the
SGI-ALTIX supercomputer at LRZ (Germany)  through the Extreme Computing
Project (DECI) SIMU-LU. 

JEFR thanks Thierry Sousbie for releasing his {\tt Skeleton} code from which 
the present work was built on, and Noam Libeskind for a careful reading of the
manuscript. SG acknowledges a Lady  Davis  Fellowship at the Hebrew University
Jerusalem. AK acknowledges support of NSF grants to NMSU. GY would like to thank  MEC (Spain) for financial support under project
numbers FPA2006-01105 and AYA2006-15492-C03.

\appendix
\section[]{Setting the threshold for web classification}
\label{app:setting}

In this paper we use the eigenvalues of the deformation tensor,
Eq.(\ref{eq:hessian}), normalized in a specific way. Here we provide details
of the normalization and give motivation for selecting the threshold.

We write the Poisson equation in the following form:  
\begin{equation}
\label{eq:poisson}
\nabla^2 \tilde{\phi} = 4 \pi G \bar{\rho} \delta = \tilde{\lambda}_1 +
\tilde{\lambda}_2 +  \tilde{\lambda}_3, 
\end{equation}
\noindent
where $\bar{\rho}$ is the mean matter density of the universe and $\delta$ is the
matter overdensity. One can re-scale the gravitational potential and the
eigenvalues of the deformation tensor by dividing them by $4\phi G\bar{\rho}$:
\begin{equation}
\label{eq:poisson-1}
\nabla^2 \phi = \delta  = \lambda_1 + \lambda_2 +  \lambda_3.
\end{equation}

Note that $4\phi G\bar{\rho}$ provides a natural scale to introduce
dimensionless parameters $\lambda_i$. We solve this equation numerically.

The spherical collapse model is invoked here so as to  get a rough estimate of $\lambda_\mathrm{th}$. 
The (spherical) free-fall time is related to the local density by 

\begin{equation}
\tau_{ff} = \sqrt{\frac{3\pi}{32G\rho}}.
\label{eq:tff}
\end{equation}

Recalling  that 
\begin{equation}
4\pi G\bar{\rho} = \frac{3}{2}\Omega_{m}H_{0}^2,  
\label{eq:cosmo-factor}
\end{equation}
(where $\Omega_m$ is the value of the cosmological matter density and $H_{0}$ is
the Hubble constant),  Eq.(\ref{eq:poisson}) is rewritten as:

\begin{equation}
\nabla^2  \tilde{\phi} = \tilde{\lambda}_{1} +\tilde{\lambda}_{2}
+\tilde{\lambda}_{3} = 4\pi G\rho - 4\pi G\bar{\rho}.
\label{eq:poisson-2}
\end{equation}

The threshold can be estimated by demanding that the free-fall time equals the age of the universe ($\tau_0$). Namely,   the threshold separates between the principal axes that have collapsed by $\tau_0$ and the ones that have not.
Substituting the free-fall time by the age of the universe the threshold is given by

\begin{equation}
3 \beta(\lambda_i) \tilde{\lambda}_\mathrm{th} = \frac{3\pi^2}{8\tau_0^2} - \frac{3}{2}\Omega_m H_0^2,
\end{equation}

where $\beta$ is a  fiducial factor introduced to account for the deviation from local isotropy.

In terms of the dimensionless eigenvalues the threshold is given by:
\begin{equation}
  \lambda_\mathrm{th} = \frac{1}{3
  \beta(\lambda_{i})}\left[\frac{\pi^2}{4}\frac{1}{\Omega_m} (\tau_0 H_0)^{-2} - 1\right]
\label{eq:final}
\end{equation}

For the WMAP3 parameters used in the $160\hMpc$ simulation we have $\Omega_m = 0.24$,
$h=0.73$ and $\tau_0 H_0 = 0.983$. It follows that $\lambda_\mathrm{th} =
9.63/(3.0\beta)$.

Assuming the spherical collapse model as a a proxy to the full non-isotropic case,  
namely   $\beta(\lambda_i)=1$, then the threshold is  $\lambda_{th} = 3.21$.

\bibliographystyle{mn2e}

\end{document}